\documentclass[review]{elsarticle}

\usepackage[]{units}
\usepackage{amsmath}
\usepackage{amsfonts}
\usepackage{amssymb}%
\usepackage{graphicx}% Include figure files
\usepackage{float}
\usepackage{dcolumn}% Align table columns on decimal point
\usepackage{bm}% bold math
\usepackage{physics}
\usepackage{lineno,hyperref}
\usepackage[dvips]{color}

\begin{document}

\begin{frontmatter}

\title{On the Sagnac effect and Quantum Mechanics  in a Rotating Reference System}

\author{Bj\o rn Jensen}

\ead{bjorn.jensen@usn.no}

\address{Department of Microsystems, \\
University College of Southeast Norway, 3603 Kongsberg, Norway}

\begin{abstract}
We canonically quantize a spin-less non-relativistic point particle in a rigidly rotating cylinder symmetric reference system. The resulting quantum mechanics is investigated in the case of both a two-dimensional cylindrical shell and in the case of the full rotating cylinder; in both cases energy eigenstates exist which exhibit rotation induced negative energies. Based on a reanalysis of the classical Sagnac effect a novel way to compute the quantum Sagnac phase shift is deduced. It is pointed out that certain states on both the cylindrical shell and in the bulk give rise to a novel anomalous quantum Sagnac effect.
\end{abstract}


\end{frontmatter}

%\linenumbers

\definecolor{brown}{rgb}{0,1,0}

%\maketitle

\section{Introduction}

Quantum field theory and quantum mechanics in both relativistic and non-relativistic rotating reference systems have recently been shown growing interest. This is partially due to fundamental issues posed by such systems in the very formulation of quantum theory (see e.g. \cite{Leinaas}), and partially due to progress in neutron interferometry and micro- and nano-technology (see e.g. \cite{Rauch}). This work deals with the non-relativistic quantum mechanics of a spin-less point particle in a non-relativistic rigidly rotating cylinder symmetric reference system.   

Part of the reason for the growing interest in quantum physics in non-relativistic rotating reference systems stems from the realization that an effect which resembles the Sagnac effect \cite{Sagnac} exists as a genuine quantum effect \cite{Werner}. It has been shown experimentally that when two beams composed of relatively slow moving quantum particles (e.g. neutrons) are emitted on a rotating platform and made to interfere after having traveled in opposite directions about the center of rotation the particles develop a relative rotation dependent phase which resembles the one appearing in the Sagnac effect \cite{Werner}. In the literature this effect is referred to as the {\it quantum Sagnac effect}\footnote{This term was probably introduced in \cite{Anandan_term}. The term is in general tied to quantum physics in both relativistic and non-relativistic rotating reference systems.}. The Sagnac effect  is central in e.g. electrodynamics in a rotating reference system where it has found important applications in laser based gyroscopes \cite{Macek} in a wide range of different application areas; e.g. in aviation and car industry and in oil exploration \cite{Anderson}. It is likewise believed that the quantum Sagnac effect can be used to construct solid state gyroscopes. Implemented as so-called atom chips they are expected to exhibit extremely high precision compared to the laser based gyroscopes \cite{Search}. However, in contrast to the Sagnac effect, and in part due to unresolved fundamental issues, the current theoretical understanding of the quantum Sagnac effect leaves a lot to be desired. 

The quantum Sagnac effect is often discussed on the semiclassical level or through {\it assumptions} about the form of the wave-function of a quantum particle in a rotating reference system. In the latter case considerations involving Galilean boost transformations (e.g. \cite{Mashhon,Suzuki}) and a simple coordinate transformation of the Sch\"{o}dinger equation from the Cartesian coordinate system to a rotating one (e.g. p.272 in \cite{Rauch} ) are common as well as arguments based on formal analogies between quantum mechanics in rotating coordinates and quantum mechanics coupled to Maxwell's electrodynamics (e.g. \cite{AharonovC,Aharonov,Sakurai_1980}). Elaborations involving the non-relativistic limit of the Dirac equation do also exist (e.g. \cite{Hendriks}). Interestingly, fundamentally divergent views on how to construct a proper rigorous formulation of non-relativistic quantum mechanics in a rotating reference system are also expressed in the literature \cite{Mashhon, Suzuki,Anandan, Mashhon_2}. The current situation concerning the theoretical treatment of the quantum Sagnac effect is thus that {\it no general and consistent and generally accepted framework based on "first principles"  exists in the literature for a non-relativistic quantum mechanics in a non-relativistic rotating reference system}\footnote{A similar statement can consequently be made about the quantum Sagnac effect in non-relativistic quantum mechanics. This author is not the first one to make such an observation concerning the status of the theoretical understanding of the Sagnac effect in quantum mechanics. A similar statement can be found about the quantum Sagnac effect in relativistic quantum mechanics in \cite{Rizzi}: "{\it ... a clear and universally shared derivation for matter waves is not available as far as we know,...}" (\cite{Rizzi} p. 183).}. The ambition of the present work is to remedy this situation; to construct a  "constructive" non-relativistic quantum mechanics for a spin-less point particle in a non-relativistic rigidly rotating reference system based on "first principles". 

The rest of this paper is organized along the following lines. In the next section we provide a discussion and a reanalysis of the Sagnac effect. It is noted that besides the special theory of relativity a certain reflection transformation does also play a central role in the description of the Sagnac effect in classical physics. In Sec. 3 we discuss the quantum Sagnac effect. It is emphasized that the reflection transformation noted in Sec. 2 is expected to be of key importance for the theoretical description of the quantum Sagnac effect. In Sec. 4 we quantize the free spin-less point particle in a non-relativistic rigidly rotating cylinder symmetric reference system when the rotation axis coincides with the axis of cylinder symmetry. In Sec. 5 we study the quantum mechanics we deduced in Sec. 4 dimensionally reduced on to a "thin" rotating straight cylindrical shell; because the resulting theory is exactly solvable, and thus able to elucidate certain aspects of the correspondning three-dimensional quantum mechanics, and because it is thought to have some relevance to quantum physics on thin micro- and nano-sized cylindrical rotating shells. In this theory we also add a magnetic flux line confined to the axis of rotation in order to see explicitly both the interplay and similarities between rotation and magnetic flux in the quantum mechanics we have constructed. One general feature of the dimensionally reduced theory is that rotation induced {\it negative energy eigenstates seems generic}.  In this context we also provide an {\it independent} calculation of the quantum Sagnac effect. It is noted that it is possible theoretically to construct wave-packets which, when made to interfere, will give rise to the quantum Sagnac effect even when they move in the {\it same} direction in space. In Sec. 6 we consider the energy eigenvalue problem in full three dimensions. This problem is not exactly solvable in general. However, in one particular instance it is. In that particular case the degeneracy of the energy of angular momentum eigenstates in the ordinary non-rotating cylinder symmetric coordinate system is lifted due to a {\it quantum angular momentum - rotation coupling}. This coupling also emerges in the dimensionally reduced quantum mechanics in Sec.5. Calculations of the energy spectra inside a micro-sized straight rotating cylinder with either the Dirichlet or the Neumann boundary condition imposed on the wave-function on the cylinder surface reveal the presence of a finite number of rotation induced {\it negative energy eigenstates} for sufficiently large angular velocities. Corresponding negative energy eigenstates are also present in the dimensionally reduced quantum mechanics in Sec. 5. Appropriately chosen exact states in three dimensions are shown to give rise to a {\it time-dependent} Sagnac effect  when made to interfere. This in stark contrast to the constant {\it time-independent} quantum Sagnac effect reported in the previous literature. In the final section we discuss our findings. The content of this paper is of interest to researches in several disparate fields. It is therefore written in a style which attempts to make the content accessible to a wider audience.

\section{The Sagnac Effect}
 
Not long after Lord Rayleigh published his study "{\it On waves propagated along the plane surface of an elastic solid}" \cite{Lord} in 1885, and well before Sagnac's seminal work \cite{Sagnac} appeared in 1913, a Sagnac effect was reported observed in 1890 when a vibrating champagne glass was set into rotation giving rise to audible beat tones \cite{Bryan}. This so called surface acoustic Sagnac effect has recently been implemented in MEMS\footnote{Micro-Electro-Mechanical-Systems.}-devices in the form of SAW\footnote{Surface Acoustic Wave, which refers to a Rayleigh wave in the theory of elasticity.} gyroscopes (see e.g. the review in \cite{Xia}).  A proposal for a corresponding MEMS-design based on ultrasound in air-ducts also exists \cite{Yu}. One might naively expect that classical Newtonian physics complemented with a wave-theory for light is capable of describing the Sagnac effect in all classical systems such as e.g. the ones referred to above as well as those based on light. However, this is not the case. A classical theory which is framed within the Newtonian perspective on space and time and which is also {\it fully} Galilean invariant in the sense that physics and all measurable parameters are the same for a rotating observer and a non-rotating observer does {\it not} in general predict an effect which corresponds to the Sagnac effect for light (see e.g. \cite{Dieks,Dieks2}). A proper understanding of the Sagnac effect in classical systems requires the special theory of relativity\footnote{As a historical side note which might arguably be taken as an illustration of the inherit problem with explaining the Sagnac effect within non-relativistic physics is to observe that M.G. Sagnac in \cite{Sagnac}, and hence {\it after} the advent of the special theory of relativity, took the observed interference pattern which today bears his name as a direct {\it proof} for the existence of a luminiferous aether.}. 

Within the framework of the special theory of relativity it is rather straightforward to show that {\it any} type of classical signal, irrespective of its source and constitution, will in complete analogy with the Sagnac effect for light give rise to a corresponding Sagnac effect. The experimental setup used by Sagnac in his discovery of the Sagnac effect is today known as a Sagnac interferometer. It basically directs two light beams emitted from the same source which is attached to a rotating platform in opposite directions along the same {\it closed} path around the center of the rotating platform. After completing one (or any number of) {\it full} roundtrip(-s) the beams are made to interfere. An interference pattern which depends linearly on the angular velocity of the platform can then be observed. How the Sagnac effect for light, as well as for any other classical means of transmitting signals, arises in a Sagnac type interferometer on the theoretical level can be shown in the following way. In four space-time dimensions the Born metric, which is the metric naturally associated with an observer moving with a constant angular velocity about an axis, can be written in the form \cite{Langevin,Rosen}
\begin{equation}
ds^2=-c^2dT^2+dr^2+\gamma^{-2}r^2d\varphi^2+dz^2\, .
\end{equation}
In this expression $z$ denotes the coordinate in the axial direction and $(r,\varphi )$ denotes the polar coordinates in the {\it rotating} frame. We have also defined
\begin{equation}
dT\equiv c\gamma dt+\frac{\Omega r^2}{c\gamma}d\varphi\, .
\end{equation}
In this expression $t$ represents the global time coordinate in the LAB-frame (i.e. the corresponding cylinder symmetric frame used by a {\it non-rotating} observer) and the rotating frame, $\Omega$ denotes the angular velocity of the rotating frame relative to the LAB-frame, $c$ the speed of light and
\begin{equation}
\gamma\equiv \sqrt{1-\frac{\Omega^2r^2}{c^2}}\, .
\end{equation}
$T$ is naturally interpreted as the local {\it proper} time coordinate relative to the rotating observer \cite{Rosen}. We also assume $r<c/\Omega$ in order to have a well defined metric. Note that $dT$ is not a total differential in general; $T$ expressed as a function of $t$ and $\varphi$ can therefore not  be derived through integration in general. However, it is a total differential if we confine our attention to a rotating circle ${\cal S}$. Let us therefore for simplicity and transparency first consider rotating observers using standardized clocks on ${\cal S}$. We assume that the circle has radius $r$ and that it is centered at the origin in the cylindrical polar coordinate system in the LAB-frame. A {\it local} proper time coordinate $T$ on the circle can then be deduced in terms of $t$ and $\varphi$. From Eq.(2) it follows that it is, apart from an additive constant which we will discard in the following, given by
\begin{equation}
T(t,r,\varphi )=c\gamma \int_{t_0}^{t} dt+\dfrac{\Omega r^2}{c\gamma}\int_{\varphi_0}^{\varphi} d\varphi \, ,
\end{equation}
where $t_0$ and $\varphi_0$ are two in general arbitrary constants. Consider two identical {\it classical} signals of {\it any} kind which are being directed in {\it opposite directions} along ${\cal S}$ from a {\it co-rotating} emitter on ${\cal S}$. Let $(t_i,T_i)$ denote the time intervals relative to the global and local time measures needed for signals to transverse the full circle in the clockwise direction $(t_1, T_1)$ and the counter clockwise direction $(t_2,T_2)$ respectively. We assume that both signals are emitted at $t=0$ relative to the LAB-frame and at $T=0$ relative to the emitter. From Eq.(4) it then follows that
\begin{eqnarray}
T_1&=& c\gamma t_1+\frac{\Omega r^2}{c\gamma}\int_0^{2\pi}d\varphi =
c \gamma t_1+\dfrac{2\pi\Omega r^2}{c\gamma}\, , \\
T_2&=& c\gamma t_2+\frac{\Omega r^2}{c\gamma}\int_0^{-2\pi}d\varphi =
c \gamma t_2-\dfrac{2\pi\Omega r^2}{c\gamma} \, .
\end{eqnarray} 
Since the metric in Eq.(1) does not differentiate between the {\it directions} the signals transverse the circle the signals will have the same (proper) speed relative to the rotating system of coordinates. It then follows by necessity that $T_1=T_2$. Hence,
\begin{equation}
t_2-t_1\equiv \Delta t=\dfrac{4\pi\Omega r^2}{c^2\gamma^2}\, .
\end{equation}
This result is clearly independent of wether signals in vacuum or signals in a material medium are considered as long as the medium treats propagation in the two directions on an equal footing. Apparently, an non-rotating stationary observer will in general register {\it different} arrival times for counter propagating signals relative to the global time-coordinate $t$. This conclusion is easily verified theoretically for the case of light signals in vacuum by considering null-geodesics (e.g. by studying the condition $ds^2=0$) on ${\cal S}$. From the expression in Eq.(7) it follows that we to leading orders in $c^{-2}$ have
\begin{equation}
\Delta t\sim \dfrac{4\pi\Omega}{c^2} r^2 \left[1+\left(\dfrac{\Omega r}{c}\right)^2+\dots \right] .
\end{equation}
For signals in the form of light rays in vacuum with a specified frequency $\nu$ we thus arrive at the celebrated expression for the Sagnac phase-shift $\Delta\phi\equiv 2\pi\nu\Delta t$ by only retaining the leading term in the expression in Eq.(8)\footnote{Note that this simple relation between phase- and time-differences does not hold in general for light waves in a material medium. E.g. due to refraction properties of light in a rotating ring of glass the discussion of the associated phase-shift becomes more elaborate than in vacuum but the time-difference between signals to first order in $\Omega$ turns out to be the same as in Eq.(7) \cite{Arditty}. In e.g. \cite{Vugalter} a similar discussion can be found for magnetostatic- and SAW-waves; the phase difference between such counter propagating waves only depends on the wave frequency and the angular velocity of the interferometer.} 
\begin{equation}
\Delta t\sim \dfrac{4}{c^2}\Omega\pi r^2\equiv \frac{4}{c^2}\Omega {\cal A}\, .
\end{equation}
We emphasize that we {\it nowhere} in the analysis leading to Eq.(7) used the assumption of dealing with light-rays or any other {\it particular kind} of classical signaling system; only properties of the {\it geometry} of space-time have been applied together with the assumption that the vacuum or the material medium in which the signals are propagating transmits the signals with the same (proper) speed in the two opposite directions of propagation. This treatment thus entails a certain {\it universality} to the Sagnac effect in {\it classical} physics. 

From the discussion above it follows that the Sagnac effect is at least {\it in part} about that {\it local} co-rotating observers positioned along the full circle are not able to mesh together their local reference systems so as to define a {\it global} concept of simultaneity on ${\cal S}$ in terms of local proper clocks since $T(t,r,0)\neq T(t,r,2\pi )$. When considering actual signals the Sagnac effect is {\it phenomenologically} about that counter propagating signals which spend the {\it same} amount of time in transition between an emitter and a receiver relative to co-moving clocks which measure proper time spend {\it unequal} amounts of time in transition relative to the global time coordinate $t$, which is the canonical time-measure in the LAB-frame. The Sagnac effect thus apparently represents a vivid demonstration of the {\it relativity of simultaneity}. 

The line of reasoning with the rotating circle above goes through in full four dimensional space-time with signals moving on arbitrary closed paths ${\cal J}$ encircling the origin; the same expression for the Sagnac phase-shift as the one we deduced above results. This is easily seen from the expressions above by assuming $r$ to be an arbitrary periodic function of $\varphi$ (and in general $z$) since all integrands depend on the {\it square} of $r$. Contributions stemming from parts of the paths which deviate from an appropriately defined perfect circle which defines the same minimum area as ${\cal J}$ when integrating $dT$ will therefore cancel exactly. It is tempting to suspect that this goes a long way to indicate that acceleration is not needed for the Sagnac effect to occur as one otherwise might think simply on the basis of the experimental setup of a Sagnac interferometer since the effect only requires a single space dimension for its description, as our treatment of the rotating circle also suggests. Indeed, one can start with a $(1+1)$-dimensional Minkowski space-time with space coordinates differing by a fixed value identified and then simply consider a simple Lorentz-boost. A multivalued time-coordinate in the boosted system of coordinates results. This implies a similar problem in defining a sense of global simultaneity in the boosted reference system as to the one on ${\cal S}$. That acceleration seems non-essential in the Sagnac effect has apparently been verified by a recent experiment through a "{\it Modified Sagnac experiment for measuring travel-time difference between counter-propagating light beams in a uniformly moving fiber}"  \cite{Wang} (a theoretical analysis of these findings can be found in \cite{Angelo}). This indicates that {\it acceleration} is indeed {\it not} essential for the manifestation of the Sagnac effect. The manifestation of the effect apparently only requires the light source to be in {\it uniform motion}. The Sagnac effect can consequently apparently be conceptualized in terms of the elementary application of the Lorentz-boost transformation in $(1+1)$-dimensions. In this sense the Sagnac effect can apparently be seen as a purely (global) {\it kinematical} effect. 

From the discussion above one might be lead to view the impossibility of defining a notion of global simultaneity on ${\cal S}$ as in some sense {\it the} source for the Sagnac effect. However, a lack of a global notion of simultaneity would not naively be expected to be manifested {\it physically} in the interference pattern generated by two beams of light but only to represent a theoretical "artifact" which emerges when trying to force {\it local} clock synchronization to a {\it global} one. This view seems to be supported by the fact that $\Delta t$ can be shown to be insensitive to the choice of synchronization method of the rotating clocks \cite{Rizzi_Serafini}. Hence, arguably something more in addition to relativity seems to be at work in the Sagnac effect. We emphasize that even though the lack of a notion of global simultaneity seems {\it integral} and {\it necessary} to the deduction of a Sagnac effect it is mathematically not strictly {\it sufficient} for having a Sagnac effect. The $T$-coordinate in Eq.(4) could, if stemming from another hypothetical space-time kinematics or clock synchronization method, in principle have depended on the angular coordinate in a different manner, like e.g. $\sim\varphi^2$. Such an angular dependence still gives rise to a multi-valued $T$-coordinate, and thus still implies the problem with meshing together the local reference systems along the full circle when attempting  to define a consistent notion of global simultaneity on ${\cal S}$ in terms of $T$, but it will {\it not} imply a Sagnac effect. Besides the multi-valuedness of $T$ we thus find at the core of the derivation of the Sagnac effect in terms of the $T$-coordinate in Eq.(4) that when 
\begin{equation}
{\cal P}:\varphi\rightarrow -\varphi\, ,
\end{equation}
we have 
\begin{equation}
{\cal P}: T(t,\varphi )\rightarrow T(t,-\varphi )\neq T(t,\varphi )\, .
\end{equation}
In other words, 
\begin{center}
{\it fundamental to the Sagnac effect is that $T$ is not invariant under the reflection transformation ${\cal P}$.}
\end{center}
Since ${\cal P}$ is derived from the treatment of the metric above it follows that ${\cal P}$ is to be understood as an {\it active} transformation within the {\it same} coordinate system. This implies in particular that the orientation of the coordinate basis vectors, and thus also of $dT$ \footnote{Strictly speaking, it is the differential form $\underline{d}T$ (which represents a {\it geometrical} object) which is dual to the coordinate basis vector tangent to the world-line traced out by an observer at rest in the rotating coordinates which does not change.}, do not change under ${\cal P}$. As far as the metric above is concerned $\Omega$ is treated as an externally controlled parameter which is defined to be invariant under ${\cal P}$. From these considerations it follows that the {\it metric} is invariant under ${\cal P}$, i.e. ${\cal P}$ represents an example of an {\it isometry}. 

{\it To summarize}: we emphasize the {\it non-invariance} of $T$ under ${\cal P}$ as a {\it prerequisite} for the Sagnac effect to occur. {\it Invariance} (hypothetical) of $T$ under ${\cal P}$ does {\it not} imply a Sagnac effect. However, in {\it both} cases time intervals in terms of $T$ and the corresponding intervals in terms of $t$ will in general differ and hence imply a  relativity of simultaneity. Hence, having a {\it kinematics} of the kind represented by the special theory of relativity is {\it necessary}  for being able to account for the Sagnac effect but it is not a {\it sufficient} condition. This line of reasoning thus puts the reflection transformation ${\cal P}$ at the center of the core of the Sagnac effect. In other words, we have through ${\cal P}$ pointed to a property of the space-time {\it geometry} defined by rotating observers equipped with standardized clocks which appears to be at least as fundamental to the Sagnac effect on the theoretical level as a relativity of simultaneity. 

\section{The Quantum Sagnac Effect}

A Sagnac effect has been measured for {\it quantum} particles moving at non-relativistic velocities \cite{Werner}. One would naively think that the Schr\"{o}dinger equation represents a fully Galilean invariant theory placed within a {\it Newtonian} perspective on space and time. It is therefore intriguing \cite{Dieks} that it {\it does} imply a {\it quantum Sagnac effect} which is consistent with measured phase shifts \cite{Werner}. The reason for this capability can arguably be traced to a non-Galilean {\it feature} of the Schr\"{o}dinger equation \cite{Dieks}; the phase of the wave-function, which one often neglects on the basis that it is of no real physical significance, does not transform under a Galilean boost-transformation as one would expect from a fully Galilean invariant theory \cite{Dieks}. In order to rough out a line of reasoning leading to this conclusion about the transformation properties of the phase of the wave-function it is sufficient to resort to the rotating circle ${\cal S}$ above. However, assume first that we consider a quantum particle with mass $m_0$ which moves along the positive (stationary and inertial) $x$-axis with velocity $v$. The associated de Broglie phase $\phi_q$ is defined by $\hbar\phi_q \equiv px -Et$ where $p$ and $E$ are the momentum and the energy of the particle respectively. With $E=\frac{1}{2}m_0v^2$ and $p=m_0v$ we get
\begin{equation}
\phi_q =\frac{1}{\hbar}(m_0vx-\frac{1}{2}m_0v^2t)
\end{equation}
as the expression for the de Broglie phase relative to the non-rotating laboratory frame. When the Galilean boost transformed of $\phi_q$ is applied {\it as is} in the experimental set-up we considered for the Sagnac effect on the rotating circle above (by identifying $x$-positions differing by a fixed value), and with the same general assumptions applied but with {\it quantum particles} traveling with the same speeds in opposite directions on ${\cal S}$ and made to interfere after each particle has traveled a {\it full} circle, the quantum Sagnac phase-shift $\Delta\phi_q$ follows \footnote{Note that in the experiments reported in the seminal paper \cite{Werner} the particle beams were made to interfere after having travelled {\it halfway} around the apparatus; the reported measured phase difference was therefore one half of the one in Eq.(13).} \cite{Anandan_term,Dieks}
\begin{equation}
\Delta\phi_q =\frac{4m_0}{\hbar}\Omega{\cal A}\, .
\end{equation}
The phase-shift in Eq.(13) represents an expression for the impossibility to define a globally well defined phase in {\it all} co-moving frames along a rotating circle ${\cal S}$ at the same {\it time} \cite{Dieks}, i.e. simultaneously relative to the Newtonian time-coordinate $t$. As is pointed out by many authors (e.g. \cite{Anandan_term,Dieks}) this behavior bears strong resemblance to the behavior of the  phase of an electrically charged particle in the Aharanov-Bohm effect \cite{AB}. 

The expression for $\Delta\phi_q$ in Eq.(13) may at first sight look very different from the corresponding expression for the phase difference  $\Delta\phi$, which can be pieced together from the results above, generated by classical signals. However, the expressions for $\Delta\phi$ and $\Delta\phi_q$ can be brought onto a similar form by evoking {\it wave-particle duality} to re-express $\Delta\phi_q$ in terms of the speed $v$ of the quantum particles. Wave-particle duality implies $\lambda_Bm_0v=2\pi\hbar$ where $\lambda_B$ denotes the Broglie wavelength of a particle. This expression can be used to "remove" $m_0/\hbar$ from the expression for $\Delta\phi_q$ such that we get
\begin{equation}
\Delta\phi_q=\frac{8\pi\Omega{\cal A}}{\lambda_Bv}\, .
\end{equation}
In this form connection to the corresponding expression for the classical Sagnac phase difference $\Delta\phi$ can be made, at least on a formal level, by setting $v\equiv c$ and $\lambda_B\equiv c\nu^{-1}$. Then the formal expressions for $\Delta\phi_q$ and $\Delta\phi$ coincide. This formal similarity may be taken as a "hint" that the issue of Einstein synchronization of clocks relative to a rotating observer is also at work in the quantum Sagnac effect; that $\phi_q$ can not be uniquely specified everywhere in a rotating system thus arguably implies that Schr\"{o}dinger theory can be seen as a "{\it non-relativistic approximation scheme to a relativistic, Lorentz invariant theory.}" \cite{Dieks}. From these considerations one might feel tempted to conclude that not only can the quantum Sagnac effect and the Sagnac effect be put on an equal footing but also Sagnac effects for massive and massless particles. 

Clearly, on general grounds {\it some} (guiding) principles must be employed when confronted with the task of bringing non-relativistic quantum mechanics into the realms of relativity theory and accelerated reference systems; Galilean boost transformations and wave-particle duality are two possible alternatives which have been studied in the literature. We want to direct the attention of the reader to one guiding principle which has not been explored in our context in the previous literature. We should without having to appeal to wave-particle duality, and a line of reasoning which might appear rather contrived, or without having to appeal to certain transformation properties of the Schr\"{o}dinger equation expect that the Sagnac effect and the quantum Sagnac effect are intimately related on the basis of  the {\it correspondence principle; we should always expect to regain the classical description of the Sagnac effect from a proper quantum mechanical formulation of the effect in the (classical) limit of large quantum numbers for appropriately chosen quantum states}. Due to the correspondence principle and the importance of ${\cal P}$ we demonstrated on the classical level we would in particular {\it expect} that ${\cal P}$ also plays a similarly important role in non-relativistic quantum mechanics in a non-relativistic rotating reference system.  We will see below that this is indeed the case.

In much of the available literature the boost approach to defining quantum mechanics in a rotating system is the preferred one. However, this approach raises several issues. {\it Firstly}, the most serious issue with this approach is that it implicitly relies on the fundamental assumption that there is a (Galilean) boost relation between quantum states defined in the stationary non-rotating frame and {\it all} quantum states in the rotating frame. However, there is {\it a priory} no guaranty that this is the case. E.g. energy eigenstates could in principle be definable in the rotating frame which does not correspond through a boost transformation to energy eigenstates in the non-rotating one, and vice versa. Hence, in the boost approach to deduce quantum states in the rotating frame one obviously runs the risk to miss the existence of novel quantum states. {\it Secondly}, it is often tactically {\it assumed} that the boost transformed of the wave-function can be applied in any one of the instantaneous inertial reference frames constructed at all points on the paths of the counter rotating particles. In the one-dimensional space defined by ${\cal S}$ this approach is probably unproblematic. However, as demonstrated by the discussions in \cite{Mashhon, Anandan, Mashhon_2,Suzuki}, and in contrast to the corresponding situation in {\it classical mechanics}, it is not unproblematic in the general situation in two or higher space dimensions to {\it assume} that it is correct to express {\it quantum mechanics} in a {\it non-inertial} frame by insisting that quantum mechanics in an {\it accelerated} frame can be derived by equating it with a formulation in the instantaneous {\it inertial} frame coinciding with the local instantaneous rest frame of the accelerated frame at a point. {\it Thirdly},  by making the non-rotating reference system essential in the deduction of the quantum Sagnac effect one might get the impression that the effect is a {\it fictitious} one and not a physical effect which is {\it intrinsic} to quantum mechanics in the rotating frame \footnote{We note that in \cite{Dieks} this issue is rather pronounced since wave-packets which are defined relative to the {\it non-rotating} frame are used in arguing for the presence of the quantum Sagnac effect.}. The boost approach thus makes the subject rather opaque; from a conceptual point of view a more transparent treatment of the subject should therefore be sought for. 

In this paper we will seek to avoid the issues noted above in constructing a non-relativistic quantum mechanics in a rotating reference system through application of Dirac's quantization program \cite{Dirac}; through direct canonical quantization of the classical mechanics of a (initially) non-interacting point particle as it is formulated in the rotating reference system. We are then in position, at least in principle and in contrast to the boost approach,  to {\it calculate} all possible quantum states in the rotating system. In this way we furthermore simply {\it bypass} previous discussions as to {\it how} quantum mechanics in a non-relativistic rotating reference system is to be formulated. We do then in particular not need to include a non-rotating reference system {\it explicitly} in the discussion of the quantum Sagnac effect, nor do we need to disentangle the nature of the transformation properties of the phase of the wave-function when changing the description of quantum mechanics from one reference system to another one either.  In our approach 
\begin{center}
{\it the quantum Sagnac effect emerges as a direct consequence of the non-invariance of the \\quantum particle Hamiltonian H under }${\cal P}$.
\end{center}
The {\it non-invariance} of the {\it quantum time-translation operator} $H$ under ${\cal P}$ allows the quantum Sagnac effect to be perceived as an effect on par with the Sagnac effect, which, as we have seen, (in part) arises due to the {\it non-invariance} of the {\it proper time-coordinate T} under ${\cal P}$. This seems to strengthen the view that the Sagnac and the quantum Sagnac effects are indeed intimately connected since the {\it corresponding} non-invariance of the local {\it time-coordinate} $T$ and the quantum {\it time-translation} operator $H$ under ${\cal P}$ seems to support the {\it correspondence principle} in quantum mechanics in our context. The correspondence between the behavior of $T$ and $H$ under ${\cal P}$  arguably implies in some sense the presence of the issue of clock synchronization in a rotating system also on the level of non-relativistic  quantum mechanics. This correspondent behavior thus represents a key ingredient in the basis of a framework which complements the one in \cite{Dieks} in showing how Schr\"{o}dinger theory can accommodate a result which naturally belongs to the realm of relativistic Lorentz invariant theory, but one which is  based on first principles. However, as will be shown below, perhaps even more important than being able to contribute in the carving out of a more consistent and unified picture of the Sagnac and the quantum Sagnac effects is that the canonical quantization program in our context points to the existence of fundamental quantum energy eigenstates in the rotating frame which can {\it not} be simply connected through a boost transformation to corresponding fundamental quantum energy eigenstates defined relative to the corresponding non-rotating reference frame. Hence, the framework which we develop in this paper thus not only complements an existing framework, it supersedes it. 

\section{Canonical Quantization}

In this work we will be concerned with non-relativistic quantum mechanics in a {\it non-relativistic} rigidly rotating cylindrical polar coordinate system when the axis of rotation coincides with the symmetry axis of the metric structure. In the {\it rotating coordinates} $(r,\varphi ,z)$ the metric structure is taken to be the usual Euclidean one, i.e. the line-element is taken to be
\begin{equation}
ds^2=dr^2+r^2d\varphi^2+dz^2\, ,
\end{equation}
with $r\geq 0$ and $-\infty < z<+\infty$. We will find it useful not to restrict the values on the angular coordinate, i.e. we will let $-\infty <\varphi <+\infty$ unless confusion follows and a necessary restriction on the values of $\varphi$ is needed. Of course, points in space differing by values of the angular coordinate equal $2\pi$ are identified. In the following we will refer to this identification as the {\it fundamental identification in the angular direction}, or more simply as the {\it fundamental identification}.  $t$ will denote the {\it Newtonian} time-coordinate in the following. 

In the rest of this section we will engage the problem of devising a non-relativistic quantum mechanics for a non-interacting classical "point" particle which eventually is made to interact with {\it classical} electrical and magnetic fields in the rotating system above. In this endeavor we choose to follow Dirac's prescription for quantization appropriately adapted to curvilinear coordinates. Even though we are dealing with a rotating coordinate system we do not need to modify Dirac's quantization procedure since the metric structure is formally identical to the one in ordinary static three dimensional space. Our starting point is therefore the Lagrangian for a free classical point particle in the {\it rotating} coordinate system above \cite{Landau}
\begin{equation}
L=\frac{1}{2}m_0\vec{v}^2+m_0\vec{v}\cdot\vec{\Omega}\times\vec{r}+\frac{1}{2}m_0(\vec{\Omega}\times\vec{r})^2\, .
\end{equation}
Here $m_0$ denotes the rest mass of the particle, $\vec{r}$ its position vector, $\vec{v}$ the associated linear velocity and $\vec{\Omega}$ the angular velocity of the rotating coordinate system relative to the LAB-frame. The second term in the Lagrangian gives rise to the Coriolis "force", the third term to the centrifugal "force". In order to construct the canonical Hamiltonian operator $H$ associated with this Lagrangian we need the classical Hamiltonian $h$ expressed in terms of $\vec{v}$ and the canonical momentum  $\vec{P}$. This momentum is easily determined
\begin{equation}
\vec{P} =\frac{\partial L}{\partial \vec{v}}=m_0\vec{v}+m_0\vec{\Omega}\times\vec{r}\, .
\end{equation}
The classical Hamiltonian $h$ can then be rewritten in the form
\begin{eqnarray}
&&h(\vec{P} ,\vec{r})=\vec{P}\cdot\vec{v}-L
%&&=\frac{1}{2m_0}(\vec{P}-m_0\vec{\Omega}\times\vec{r})^2-\frac{1}{2}m_0(\vec{\Omega}\times\vec{r})^2=\nonumber\\
=\frac{1}{2m_0}\vec{P}^2-\vec{P}\cdot\vec{\Omega}\times\vec{r}\, .
%&&=\frac{1}{2m_0}\vec{P}^2-\frac{1}{2}(\vec{P}\cdot\vec{\Omega}\times\vec{r}+\vec{\Omega}\times\vec{r}\cdot\vec{P})\, .
\end{eqnarray}
Relative to the orthonormal coordinate basis vectors $\{\vec{e}_{\hat{r}},\vec{e}_{\hat{\varphi}},\vec{e}_{\hat{z}}\}$ we have $\vec{\Omega}=\Omega\vec{e}_{\hat{z}}$ such that
\begin{equation}
\vec{\Omega}\times\vec{r}=\Omega r\vec{e}_{\hat{\varphi}}\, .
\end{equation}
Relative to this basis the components of the canonical momentum are given by
\begin{equation}
\left\{\begin{array}{l}
P^{\hat{r}}=m_0\dot{r}\, ,\\
P^{\hat{\varphi}}=m_0r(\dot{\varphi}+\Omega )\, ,\\
P^{\hat{z}}=m_0\dot{z}\, ,
\end{array}\right.
\end{equation} 
where the dot notation represents differentiation with respect to $t$. We note that in a typical semi-classical approach to the quantum Sagnac effect the induced phase-shift $\Delta\phi_q$ is basically identified with the circulation of the total canonical momentum around the paths which are traced out by the particles. The phase-shift is thus defined as
\begin{equation}
\Delta\phi_q\equiv\frac{1}{\hbar}(\oint_1\vec{P}_1\cdot d\vec{r}_1+\oint_2\vec{P}_2\cdot d\vec{r}_2)\, ,
\end{equation}
where $\vec{P}_1$ and $\vec{P}_2$ denote the respective canonical momenta associated with particle $1$ and particle $2$. The integrals are computed along the respective particles trajectories. Restricting attention to particles moving in opposite directions on a circle with radius $R_0$ we easily get
\begin{equation}
\Delta\phi_q=\frac{4m_0}{\hbar}\Omega\pi R_0^2 \equiv \frac{4m_0}{\hbar}\Omega{\cal A}\, .
\end{equation}

In the following we will denote the coordinates by $q^n$ in the usual fashion such that $\{ q^n\} =\{ r,\phi ,z\}$. From the expressions for the components of the canonical momentum it follows that the determinant of the matrix with components
\begin{equation}
\dfrac{\partial^2L}{\partial\dot{q}^n\partial\dot{q}^{n'}}
\end{equation}
is non-zero. This implies that the theory does not harbor any, in the language of Bergmann and Dirac, primary constraints  \cite{Bergmann,Dirac}. Hence, the components of the canonical momentum can be treated as independent and thus form the basis for the canonical quantization procedure \cite{Teitelboim}. We quantize the theory by promoting $\{ q^n\}$ and the conjugate momenta $\{P^n\}$ to operators. The conjugate momenta are identified with the components of the gradient operator $\vec{\Pi}=-i\hbar \nabla$. In order to construct an Hermitian operator which obeys the usual commutation relations with our curvilinear coordinates $\{ q^n\}$ we follow \cite{Gruber,Leaf} and re-write the gradient in a symmetric form. Let $\Pi_n$ represent a covariant component of the resulting object in the {\it non-normalized  coordinate basis} $\{ \vec{e}_n\}$. We then define $\Pi_n\equiv 1/2 (\vec{e}_n\cdot \vec{\Pi} +\vec{\Pi}\cdot\vec{e}_n)$. It follows that relative to the coordinate basis we can write \cite{Gruber,Leaf}
\begin{equation}
\vec{\Pi} =\vec{e}_n\Pi^{n\dagger}\, ,
\end{equation}
where the Hermitian adjoint of $\Pi_n$ is given by
\begin{equation}
\Pi^{\dagger} _n=\frac{\hbar}{i}(\frac{\partial}{\partial q^n}+\frac{1}{\sqrt{g}}\frac{\partial\sqrt{g}}{\partial q^n})\, .
\end{equation}
Here $g$ denotes the determinant of the metric tensor. It follows in particular that 
\begin{equation}
[ \Pi_n,q^{n'} ]=\frac{\hbar}{i}\delta_n\,\!\! ^{n'}\, ,
\end{equation}
i.e., $\Pi_n$ is the operator which is {\it canonically conjugate} to $q^n$. We note that the metric and the fundamental commutator above are invariant under the {\it spatial translations} (separately)
\begin{equation}
\left\{\begin{array}{l}
\varphi\rightarrow\varphi +\varphi_0\, ,\\
z\rightarrow z+z_0\, ,
\end{array}\right.
\end{equation}
where $\varphi_0$ and $z_0$ denote arbitrary constants, and the {\it reflections} (separately)
\begin{equation}
\left\{\begin{array}{l}
\varphi\rightarrow -\varphi\, ,\\
z\rightarrow -z\, .
\end{array}\right.
\end{equation}
We will interpret these transformations as {\it active} transformations. The first of the reflections in Eq.(28) thus coincides with ${\cal P}$ in Sec. 2.

The expression for the classical hamiltonian $h$ symmetrized in the usual fashion with respect to the coordinates and the components of the canonical momentum provides us with the form of the associated Hamiltonian operator $H$. Due to the form of the metric Eq.(15) and the consequent resulting expressions for the components of the canonical momentum we can write the Hamiltonian operator in the following form 
\begin{equation}
H=\frac{\hbar^2}{2m_0}\vec{\Pi}^2-\vec{\Omega}\times\vec{r}\cdot\vec{\Pi}\, .
\end{equation}
Due to the structure of the second term we note in particular that 
\begin{equation}
{\cal P}: H(r,\varphi ,z)\rightarrow H(r,-\varphi ,z)\neq H(r,\varphi ,z)\, .
\end{equation}
Of course, this result also holds when both reflection transformations in Eq.(28) are applied. We will for convenience in the following only discuss the reflection transformation in the $\varphi$-direction  since all the consequent results also hold when the transformation in the $z$-direction is applied together with ${\cal P}$. It might be tempting to think that this breaking of the ${\cal P}$-symmetry is due to the Coriolis "force" on the classical level. We want to emphasize that the canonical momentum contains two terms; both contribute to the centrifugal and the Coriolis terms in the Lagrangian in Eq.(16). It is therefore not possible to ascribe the symmetry breaking to either interaction alone. Equation (30) implies that the Hamiltonian operator breaks one of the symmetries of the fundamental algebra. We emphasize that both $H$ and its ${\cal P}$-transformed represent time-translations operators on an {\it equal} footing at every point in space. Below this fact will be turned into a "mechanism" whereby the quantum Sagnac phase shift can be calculated. 

In this work we will eventually also be interested in the relation between rotation and electromagnetism on the quantum level. Coupling the quantum particle we are considering to the electromagnetic field is achieved through the usual minimal coupling mechanism
\begin{equation} 
\vec{\Pi}\rightarrow \vec{\Pi}+\frac{ie}{\hbar c}\vec{A}\, ,
\end{equation}
where $\vec{A}$ denotes the electromagnetic vector potential and $e$ the electron charge. The Hamiltonian operator can then be written in the form
\begin{eqnarray}
H&=&\frac{1}{2m_0}\vec{\Pi}^2-\vec{\Omega}\times\vec{r}\cdot(\vec{\Pi}+
\frac{ie}{\hbar c}\vec{A})-\nonumber
\\&-&
\frac{1}{2m_0}(\frac{ie}{\hbar c}\vec{A}\cdot\vec{\Pi}+\frac{e^2}{\hbar^2 c^2}\vec{A}^2)
\end{eqnarray}
when we assume that $\vec{A}\cdot\vec{\Pi}=\vec{\Pi}\cdot\vec{A}$ and $\vec{A}\cdot\vec{\Omega}\times\vec{r}=\vec{\Omega}\times\vec{r}\cdot\vec{A}$. Clearly, $[H,\Pi^i]=0$. 

In this section we have developed a (partial) quantum theory which is appropriate for the description of a classical "point" particle interacting with {\it classical} electric and magnetic fields in the rigidly rotating cylinder symmetric reference system. In addition to these elaborations we also need to investigate the structure of the physical quantum state space associated with the algebraic expressions above. This we will do through an "example" in the next section when we consider our quantum mechanics dimensionally reduced on to a rotating cylindrical shell. This study allows us to get a solid handle on the theory we develop and some of its consequences before we approach the full three-dimensional situation. In contrast to this latter case the two-dimensional model is exactly solvable in general by elementary means.

\section{Quantum Mechanics on a Thin Rotating Cylindrical Shell}

In this section we will consider the quantum theory we devised above on a straight and infinitely long cylindrical shell with zero thickness which rotates about its symmetry axis. Reducing the dimensionality of space allows an exact treatment of the energy eigenvalue problem. We can also tackle the quantum Sagnac effect exactly since based on the analysis of the behavior of the Sagnac effect above it arguably essentially only depends on only one space dimension. A rotating cylindrical shell is not only of academic interest. It is expected to capture some essential physics connected with a sufficiently long and thin rotating physical shell. 

One way to theoretically handle the dimensional reduction of a formulation of quantum mechanics in a shell with finite thickness to a shell with vanishing thickness is the technique which was developed in \cite{daCosta}. In the case of a {\it non-rotating} shell this approach to {\it deducing} (in contrast to simply declaring) a dimensionally reduced quantum theory on to a surface will result in the canonical two-dimensional Schr\"{o}dinger theory in the surface with the addition of a "geometric potential" $V_q$. This potential stems from the procedure of decoupling the dynamical degrees of freedom perpendicular to the surface from the degrees of freedom "parallell" to, and eventually in, the surface. The geometric potential depends in general on the intrinsic curvature of the surface, in the form of the Gaussian curvature, and the curvature of the surface relative to the ambient space in the form of the square of the mean curvature of the surface \cite{daCosta}. In the case of a straight cylindrical surface the Gaussian curvature vanishes. The mean curvature is inversely proportional to the square of the radius of the cylinder. The resulting geometric potential is negative and given by
\begin{equation}
V_q=-\frac{\hbar^2}{2m_0R_0^2}\, .
\end{equation} 
This approach to constructing an effective quantum theory on a surface, or similarly on a linear structure, is hampered with certain issues \cite{Jensen}. It is e.g. unclear whether the geometric potential exists in the limit of vanishing shell thickness \cite{Jensen}. Away from this limit both the intrinsic and extrinsic surface curvatures are in general expected to contribute to the particle dynamics through the boundary conditions acting on the wave-function. In this paper we will assume that the geometric potential in \cite{daCosta} represents an {\it approximation} to the physically induced potential in the case when the thickness of the surface is sufficiently small. In the case of the straight cylinder it is natural to assume that this approach returns a result which is directly physically relevant due to the high degree of symmetry of the geometry of the cylindrical surface. Certainly, we are interested in the case when the surface rotates about its symmetry axis. The procedure in \cite{daCosta} is applicable also in this case. The inclusion of rotation will not alter the geometric potential in Eq.(32). This is due to the fact that the $\vec{\Omega}\times\vec{r}\cdot\vec{P}$-term in the Hamiltonian, which is the only term in the Hamiltonian which carry the rotation parameter, only acts in the {\it angular} direction, which eventually is {\it intrinsic} to the surface, due to the form of $\vec{\Omega}\times\vec{r}$; it does therefore {\it not} contribute to $V_q$. Since the geometric potential on the rotating shell is constant we will treat it as a simple shift of the energy spectrum of  quantum particles on the surface. We will therefore neglect it in the following treatment since it will not contribute to the main objective of this work. However, we emphasize that $V_q$ is {\it negative} and will consequently imply a {\it negative} shift to all calculated energy spectra.

On the thin rotating cylindrical shell we employ the cylindrical polar coordinates $(\varphi ,z)$ such that the metric on the shell is simply given by
\begin{equation}
ds^2=R_0^2d\varphi^2+dz^2\, .
\end{equation}
$R_0$ denotes the radius of the (physical) cylinder. On the basis of the preceding paragraph we have
\begin{equation}
\vec{\Pi}=-i\hbar (\vec{e}_{\hat{\varphi}}\frac{1}{R_0}\frac{\partial}{\partial\varphi}+\vec{e}_{\hat{z}}\frac{\partial}{\partial z})\, .
\end{equation}
We will only consider time independent stationary quantum states such that the time-dependent part of the wave-function can be factored out as a simple exponential in the usual fashion. $E_+$ will denote, with quantum numbers suppressed, the energy of a state in the following. The $z$-direction can also be factorized out from the wave-function as a simple exponential with wave-number $k_+$ (with quantum numbers suppressed). The energy eigenvalue problem can consequently be written in the form
\begin{eqnarray}
\frac{\partial^2}{\partial\varphi^2}\psi_+ +Ai\frac{\partial}{\partial\varphi}\psi_+ +B_+\psi_+ =0\, ,
\end{eqnarray}
where
\begin{equation}
A\equiv -\dfrac{2m_0R_0^2\Omega}{\hbar}\,\, ,\,\, B_+\equiv -R_0^2(k_+^2-\frac{2m_0E_+}{\hbar^2})\, .
\end{equation}
The role of the the sign convention will become clear below. The time- and $z$-"directions" are "non-essential" in our context. We will therefore typically refer to $\psi_+$ as the wave-function of the particle in the following when no ambiguities are implied (and with all quantum numbers suppressed). We note that we can also derive the expression for the Schr\"{o}dinger equation in Eq.(36) through a simple coordinate transformation ${\cal Q}$ of  the corresponding  Schr\"{o}dinger equation in the {\it non-rotating} cylindrical polar coordinate system $(r',\varphi ',z')\equiv \{ q^{\prime n}\}$ with wave-function $\psi^\prime _+$ with 
\begin{equation}
{\cal Q}:\left\{\begin{array}{l}
\varphi =\varphi ' -\Omega t\, ,\\
r=r'\equiv R_0\, ,\\
z=z'\, .
\end{array}\right.
\end{equation}
It is assumed that the metric is {\it form-invariant} under the Galilean boost transformation ${\cal Q}$. The time coordinates in the two frames are also the same, of course. We emphasize that the transformation of the Schr\"{o}dinger equation in the non-rotating reference system does {\it not} involve considerations about the transformation of the {\it phase} of the wave-function $\psi^\prime _+$ under ${\cal Q}$; $\psi^\prime _+$ must be assumed to transform as an ordinary {\it scalar} quantity, i.e. $\psi_+(q^n)=\psi^\prime _+(q^{\prime n})$. This transformation rule for the wave-function contrasts with the one in e.g. \cite{Dieks}. There a rotation dependent phase factor in the transformation rule for the wave-function was also introduced. We emphasize that in our discussion the transformation properties of the wave-function stems from the properties of the {\it energy eigenvalue problem} while the discussion in \cite{Dieks} only consider the relative behavior of (certain) {\it wave-functions} in the two types of reference frames.

At this point in the development of the quantum theory we need to address the question of the {\it structure} of the physical quantum {\it state space}. We did not address this issue in the last section. We will base the construction of the state space on imposing the {\it correspondence principle} through the action of ${\cal P}$ on $H$ in the manner explained earlier in this paper. Let us therefore introduce two Hamilton operators; the operator $H_+$ defined by Eq.(36) and the operator $H_-$ defined by
\begin{equation}
H_-\equiv {\cal P}(H_+)\, .
\end{equation}
Both operators act on their respective state space with vectors we simply denote by $|\psi _-\rangle$ and $|\psi_+\rangle$ (with all quantum numbers suppressed). The time development of general physical quantum states $|\psi\rangle$ is defined to be controlled by the total Hamiltonian operator $H$ defined through (with a slight sloppiness in the notation implied)
\begin{equation}
H|\psi\rangle \equiv H_-|\psi _-\rangle + H_+|\psi_+\rangle\, .
\end{equation}
The state $|\psi\rangle$ is consequently in the position representation in general given as a sum of wave-functions on the form $\psi =\psi_++\psi_-$; each wave-function in the sum satisfies the energy eigenvalue problem in terms of $H_+$ or $H_-$. Of course, we could also have defined this structure of the state space on the level of the canonical momentum $\vec{\Pi}$ in Eq.(35) and the operator defined by ${\cal P}(\vec{\Pi})$. When $H_+$ is defined by Eq.(36) (and acting on $\psi_+$) we define through Eq.(39) the ${\cal P}$-transformed of Eq.(36) to be given by
\begin{equation}
\frac{\partial^2}{\partial\varphi^2}\psi_- -Ai\frac{\partial}{\partial\varphi}\psi_- +B_-\psi_- =0\, ,
\end{equation}
where
\begin{equation}
B_-\equiv -R_0^2(k_-^2-\frac{2m_0E_-}{\hbar^2})\, .
\end{equation}
The notation employed in these expressions should be self-explanatory. Note that there is {\it no} a priory natural relation between the {\it state spaces} spanned by the $\psi_+$- and $\psi_-$-states in terms of ${\cal P}$. The quantum state space is split into two {\it sectors}; we will refer to these as the {\bf +} and {\bf -} sectors in the following.

The general solutions of Eq.(36) and Eq.(41) can be deduced by elementary means. It turns out that {\it three} main classes of solutions exist. The {\it first} class is characterized by the condition
\begin{equation}
{\bf I:}\,\,\,  D_{\pm}\equiv A^2+4B_{\pm} > 0\, .
\end{equation}
The general solutions to Eq.(36) and Eq.(41) can in this case be written in the compact form (neglecting again for simplicity the time- and $z$-dependent factors)
\begin{eqnarray}
\psi_{\pm} &=& e^{\mp\frac{i}{2}A\varphi}(C_{2\pm}e^{\pm\frac{i}{2}\sqrt{D_{\pm}}\varphi}+C_{1\pm}e^{\mp\frac{i}{2}\sqrt{D_{\pm}}\varphi})\equiv\nonumber\\
&\equiv& e^{\mp\frac{i}{2}A\varphi}\Psi_{\pm}\, . 
%&\equiv& e^{-\frac{i}{2}A\varphi}(C_2e^{im\varphi}+C_1e^{-m\varphi})
\end{eqnarray}
As a means to accommodate an excursion to explore the relation between the current quantum mechanics and quantum mechanics in a static {\it non-rotating} frame with the interaction with an electromagnetic vector-potential $\vec{A}$ included below we choose to rewrite the last expression in Eq.(44) in the form
\begin{equation}
\psi_{\pm}=e^{-\frac{i}{2}\int_{\pm}Ad\varphi}\Psi_{\pm}\, .
\end{equation}
In this expression $\int_+$ corresponds to integration in the direction of {\it increasing} $\varphi$-values while $\int_-$ denotes integration in the opposite counter clockwise direction such that $\int_-=-\int_+$. Formally, arbitrary phases then appear. These will not have any physical significance; we therefore set these to zero in the following. We will consider the class {\bf I} solutions further in some greater detail below.

The {\it second} class of solutions of the eigenvalue problems in Eq.(36) and Eq.(41) is characterized by 
\begin{equation}
{\bf II:}\,\,\, D_{\pm}=0\, .
\end{equation}
In this case the general solutions of Eq.(36) and Eq.(41) "degenerate" into
\begin{equation}
\psi_{\pm}=e^{-\frac{i}{2}\int_{\pm}Ad\varphi}(C_{1\pm}\varphi + C_{2\pm})\, .
\end{equation}
The corresponding energy eigenvalues are given by (note the inclusion of the $z$-dependence)
\begin{eqnarray}
E_{k_{\pm}}(\Omega )&=&\frac{\hbar^2}{2m_0}(k_{\pm}^2-(\frac{m_0R_0\Omega}{\hbar})^2)\equiv\nonumber\\
&\equiv& E_{0k_{\pm}}-\frac{1}{2}m_0R_0^2\Omega^2\, .
\end{eqnarray}
Clearly, by making the $z$-contribution to the momentum sufficiently small the energy eigenvalues can be made {\it negative}, but bounded from below for a fixed value on $\Omega$.  Note that $E_{k_+}$ and $E_{k_-}$ exhibit the same form; i.e. the form of the energy eigenvalues are {\it independent} of which sector the quantum states belong to. The states in Eq.(47) represent momentum eigenstates provided that $C_{1\pm}=0$. We also note that the structure of the energy eigenvalues in Eq.(48) is that of the spectrum on a non-rotating cylindrical shell $E_{0k_{\pm}}$ (neglecting the contribution from the geometric potential) corrected by a {\it classical state independent} term.

We saw earlier in this paper that the Sagnac effect can basically be considered as a kinematical effect in a one-dimensional space. The quantum Sagnac effect should consequently emerge in our simple model with $k_\pm =0$, and indeed, it does. The states in Eq.(47) with $C_{1\pm}=0$ are particularly well suited to demonstrate this. Forming the state $\psi =\psi_++\psi_-$ we find that
\begin{eqnarray}
\hspace{-1.0cm}|\psi |^2&=&|C_{2+}|^2+|C_{2-}|^2+\nonumber\\ 
&+&C^*_{2+}C_{2-}e^{\frac{i}{2}(\int_+Ad\varphi -\int_-Ad\varphi )}+\nonumber\\
&+&C^*_{2-}C_{2+}e^{\frac{i}{2}A(\int_-Ad\varphi -\int_+Ad\varphi )}\, .
\end{eqnarray}
The interference term $\Delta\psi$ with minimum phase shift is obtained by using
\begin{equation}
\oint_+d\varphi -\oint_-d\varphi =2\oint_+d\varphi =2\oint d\varphi =4\pi\, ,
\end{equation}
such that
\begin{equation}
\Delta\psi =2\Re (C^*_{2+}C_{2-}e^{i2\pi A})\, .
\end{equation}
A Sagnac type phase shift equal $2\pi |A|$ follows. It coincides numerically with $\Delta\varphi_q$. We will therefore {\it identify} the phase shift implied by $\Delta\psi$ with the phase shift observed in experiments which involve the quantum Sagnac effect. We emphasize that this derivation of the quantum Sagnac effect only refers to quantum states defined relative to the rotating reference system. It does not at any point refer to states defined relative to the corresponding non-rotating reference system. The quantum Sagnac effect is thus seen as an effect which is {\it intrinsic} to the quantum mechanics in the rotating frame. This result is not only restricted to class {\bf II} states; it also applies to the physically more interesting class {\bf I} states (se below). We emphasize that a {\it prerequisite} for the quantum Sagnac effect to occur on the level of exact solutions to the energy eigenvalue problem is the structure of the quantum state space we introduced earlier in this section; without it {\it no} quantum Sagnac effect will emerge. To emphasize, 
\begin{center}
{\it the quantum Sagnac effect does {\it not} appear when states which belong only to either sector of the state space are made to interfere.}
\end{center}
This fact represents a strong support for the construction in Eq.(40).

The {\it third} class of solutions to Eq.(36) and Eq.(41) is defined by $D_\pm <0$. We will not study these class ${\bf III}$ solutions in this paper since that will take us too far afield.

Let us return to the class {\bf I}-solutions. The product structure of the quantum states $\psi_\pm$ in Eq.(45) are particularly interesting, for two reasons. {\it Firstly}, the structure is reminiscent with the structure of the wave function of a particle outside a thin solenoid and thus hints towards a relation to the Aharanov-Bohm effect \cite{AB}. {\it Secondly}, from a purely mathematical perspective a {\it unique} way to impose the fundamental identification in the angular direction on these states does not present itself; obviously, the fundamental identification can either be imposed on $\psi_{+}$ and $\psi_-$, {\it or} on $\Psi_{+}$ and $\Psi_-$, {\it or} on $\psi_{+}$ and $\Psi_-$ (or vice versa). Let us first consider the apparent connection between the class {\bf I} solutions and the physics of a quantum particle coupled to electromagnetism in order to shed some light on the nature of the product  structure of the $\psi_\pm$-states as it emerges in Eq.(45). We will thereafter consider the second issue.

%\subsection{On the formal relation between rotation and electromagnetism }

The factorization of the wave-function in Eq.(45), and the treatment of quantum Sagnac phase above, bears strong resemblance to the treatment of the interference of two beams of electrons passing on opposite sides of a thin magnetic flux-tube \cite{AB}. The close formal connection between a rotating reference system and electromagnetism on the quantum level has been further emphasized and probed in e.g. \cite{AharonovC}. In order to explore this relation in our context let us reformulate the quantum theory we have formulated in terms of a Lagrangian density. The most straightforward way to do this is through the Sturm-Liouville form of the equations of motion for $\psi_\pm$. Both equations Eq.(36) and Eq.(41) can be recast into the Sturm-Liouville form
\begin{equation}
\hspace{-0.219cm} e^{ i\int_{\pm}Ad\varphi}(e^{i\int_{\mp}Ad\varphi}\frac{\partial}{\partial\varphi}(e^{i\int_{\pm}Ad\varphi}\frac{\partial}{\partial\varphi}\psi_{\pm} ) -B_{\pm}\psi_{\pm} ) =0\, .
\end{equation}  
It is then straightforward to write down the associated Lagrangian density 
\begin{equation}
{\cal L}(\psi_{\pm} ,\psi^*_{\pm};\varphi )=\frac{1}{2}e^{i\int_{\pm}Ad\varphi}(\frac{\partial\psi_{\pm}}{\partial\varphi}\frac{\partial \psi^*_{\pm}}{\partial\varphi}-B_{\pm}\psi_{\pm}\psi^*_{\pm})
\end{equation}
where asterisk indicates complex conjugation. $\Psi_{\pm}$ are the natural {\it dynamical} variables in our theory. When expressed in terms of these the Lagrangian density can be written in the following form 
\begin{eqnarray}
{\cal L}(\Psi_{\pm} ,\Psi^*_{\pm};\varphi )&=&\frac{1}{2}e^{i\int_{\pm}Ad\varphi}(\frac{\partial\Psi^*_{\pm}}{\partial\varphi}\frac{\partial\Psi_{\pm}}{\partial\varphi}+\frac{i}{2}J_{\pm}+\nonumber\\
&+&(\frac{1}{4}A^2-B_{\pm})\Psi^*_{\pm}\Psi_{\pm} )\, .
\end{eqnarray}
In this exression the "current" density $J_{\pm}$ is defined by
\begin{equation}
J_{\pm}\equiv \Psi^*_{\pm}\frac{\partial\Psi_{\pm}}{\partial\varphi}-\Psi_{\pm}\frac{\partial\Psi^*_{\pm}}{\partial\varphi}\, .
\end{equation}
The Lagrangian density is structurally similar to the corresponding density for a point particle coupled to the electromagnetic vector potential through a minimal coupling mechanism in a non-rotating reference system as well as to a constant $\Omega$-dependent potential through $A^2$. We can make this similarity manifest by introducing the "covariant" derivative $D$ into the expressions above with
\begin{equation}
D\equiv \frac{\partial}{\partial\varphi}-\frac{i}{2}A\, .
\end{equation}
The Lagrangian density and the associated equation of motion for $\Psi_\pm$ can then be expressed in the following forms:
\begin{eqnarray}
\hspace{-0.25cm}{\cal L}(\Psi_{\pm} ,\Psi^*_{\pm};\varphi )=\frac{1}{2}e^{i\int_{\pm}Ad\varphi}(\overline{D}\,\Psi^*_{\pm}D\Psi_{\pm} -B_{\pm}\Psi^*_{\pm}\Psi_{\pm} ) 
\end{eqnarray}
and
\begin{eqnarray}
e^{\frac{i}{2}\int_{\pm}Ad\varphi}(\overline{D}D\Psi_{\pm} -B_{\pm}\Psi_{\pm} )=0\, ,
\end{eqnarray}
where 
\begin{equation}
\overline{D}\equiv \frac{\partial}{\partial\varphi}+\frac{i}{2}A\, .
\end{equation}
Note that even though our theory on the Lagrangian form at a superficial structural level may look like a gauge theory formulation reminiscent with the gauge theory formulation of Schr\"{o}dinger theory coupled to the electromagnetic vector potential $\vec{A}$ there is a crucial difference between them. Our theory is not gauge invariant in the usual sense. The reason on a technical level is the presence of both the $D$ and the $\overline{D}$-operators in our expressions. The Lagrangian density and the equations of motion in the gauge theory formulation of quantum mechanics coupled to electromagnetism is based solely on an operator analogous to the $D$-operator. This implies that the wave-function in that context (which we will denote by $\tilde{\psi}$ in the following) in general can be written (in full three dimensions) on the form
\begin{equation}
\tilde{\psi}(\vec{x}) =e^{\frac{ie}{\hbar c}\int_{\Gamma}\vec{A}(\vec{x'})\cdot d\vec{r'}}\tilde{\Psi}(\vec{x})\, ,
\end{equation}
where the integral can be taken along any path $\Gamma$ with endpoint at $\vec{x}$. Since
\begin{equation}
D^2\tilde{\psi}=e^{\frac{ie}{\hbar c}\int_{\Gamma}\vec{A}(\vec{x'})\cdot d\vec{r'}}\nabla^2\tilde{\Psi}\, ,
\end{equation}
where
\begin{equation}
D\equiv \nabla -\frac{ie}{\hbar c}\vec{A}\, ,
\end{equation}
it follows that $\tilde{\psi}$ will satisfy the Schr\"{o}dinger equation provided that $\tilde{\Psi}$ satisfies the Schr\"{o}dinger equation with $\vec{A}=0$. This line of reasoning does {\it not} hold in general in our formulation of quantum mechanics on the rotating cylinder. This conclusion is illustrated by the class {\bf I} solutions, but interestingly {\it not} by the class {\bf II} solutions. This latter class {\it does} have a structure which is reminiscent with the structure in Eq.(60). This last point illustrates that the relation between rotation and electromagnetism in quantum mechanics is rather complicated and not a straightforward one.

There is another important difference between our quantum mechanics and standard quantum mechanics coupled to the electromagnetic field besides the one noted in the previous pharagraph. Within the context of the latter framework $A$ will be interpreted as being an expression for the magnetic flux piercing through the center of a non-rotating cylinder (se below for details). Within that framework $\tilde{\Psi}$ will be taken as the {\it exact} quantum state when no magnetic flux is present since a gauge transformation of the wave function does not influence the energy levels of the quantum particle. In order to keep $\tilde{\psi}$ {\it single-valued} it follows that the magnetic flux {\it must} be quantized (in an analogous system constituted by a superconducting ring with a magnetic field piercing through its center this is a well known phenomenon, of course (see e.g. \cite{Tinkham})). This line of thinking conforms with experimental results. However, it is not clear how this translates to our framework. This issue is directly related to the second point we raised above; mathematically there does not present itself an unique way to impose the fundamental identification on the rotating shell. Let us consider this point in some detail.

%\subsection{Imposing the fundamental identification \\ on quantum states}

Exactly how the fundamental identification is to be imposed on quantum states on a rotating cylinder is not dictated by physics, at least not by formal similarities with quantum mechanics coupled to electromagnetism as we pointed out above. There it was emphasized that from a purely formal mathematical point of view the fundamental identification can be imposed on the quantum states in several different ways. Let us consider two important possibilities and let us still confine our attention to class ${\bf I}$ states. Let us first impose the periodicity condition implied by the fundamental identification at $\varphi =0$ and $\varphi =2\pi$ on $\Psi_+$ (the same line of reasoning will hold {\it ad verbum} for the $\Psi_-$ states, of course). This implies that 
\begin{eqnarray}
\psi_+ &\sim& e^{-\frac{i}{2}A\varphi}(C_{2m+}e^{im_+\varphi}+C_{1m+}e^{-im_+\varphi})\equiv\\
&\equiv& e^{-\frac{i}{2}A\varphi}(\psi_{m_+}+\psi_{-m_+})\, ,
\end{eqnarray}
where $m_+$ denotes integers. It follows that
\begin{equation}
\Pi^{\hat{\varphi}}\psi_+ =(m_0R_0\Omega\pm\frac{\hbar m_+}{R_0})\psi_+ 
\end{equation}
where the plus-sign in the sum in Eq.(65) follows from setting $C_{1m+}=0$ and the minus-sign from setting $C_{2m+}=0$ in Eq.(63). Note the $\hbar$-{\it independent} term in Eq.(65). The associated energy eigenvalues $E_{m_+}$ are also easily determined (with the $k_+$-momentum  in the $z$-direction included)
\begin{eqnarray}
E_{m_+}&=&\frac{\hbar^2 k_+^2}{2m_0}+\frac{\hbar^2m_+^2}{2m_0R_0^2}-\frac{1}{2}m_0R_0^2\Omega^2\equiv\nonumber\\
 &\equiv& E_{0m_+}-\frac{1}{2}m_0R_0^2\Omega^2\, .
\end{eqnarray}
These eigenvalues exhibit the usual degeneracy in the quantum numbers. The spectrum lets itself to be interpreted as the corresponding energy in the {\it non-rotating} system $E_{0m_+}$ with a state independent {\it classical} term extracted (note that this term is independent of $\hbar$ - we note the structural similarity between the spectrum in Eq.(66) and the one in Eq.(48)). A similar situation occurred in the computation of the momentum in Eq.(65) where a classical state independent term were added to the usual expression for the momentum in the corresponding non-rotating reference system. It is natural to interpret these classical terms as the momentum and energy of the co-rotating observer relative to the non-rotating inertial reference system. The same reasoning goes through for the $\psi_-$-states. Hence, states defined though periodic $\Psi_\pm$-functions represent states which in many respects resemble classical particles. If we think analogously to the way we think when a quantum particle is coupled to electromagnetism in an Aharanov-Bohm type of experiment with a thin magnetic flux tube we should not only impose periodicity of $\Psi_{\pm}$ in $\varphi$ but also insist that $E_{m_{\pm}}=E_{0m_{\pm}}$ since $\Psi_{\pm}$ is assumed to play the role of the quantum state when no rotation (i.e. correspondingly no magnetic flux) is present. Implementing this picture in our context implies that $B_{\pm}=n_{\pm}^2$ with $n_{\pm}$ denoting sets of integers. The rotation parameter then consequently gets quantized
\begin{equation}
\Omega\rightarrow \Omega_{s_{\pm}}=\frac{\hbar s_{\pm}}{2m_0R_0^2}\, ,
\end{equation}
where $s_{\pm}$ denotes other sets of integers. Clearly, this can only hold true by carefully adjusting the angular velocity artificially, or possibly through a coupling to gravity \footnote{In \cite{Anandan_1977} the Klein-Gordon equation was studied in curved space-time and the phase shift due to both rotation and curvature were discussed.}. However, no natural physical mechanism forces this quantization in our context. Hence, in the context of a rotating reference system in flat space-time $\Psi_{\pm}$ will in general {\it not} equal a corresponding state in the correspondning non-rotating reference system. This conclusion reinforces the line of reasoning based on the Lagrangian formulation above.

At this point we want to connect our discussion of the class {\bf I} solutions to the derivation of the quantum Sagnac effect in \cite{Dieks}. Consider the $\psi_{\pm}$ states in Eq.(44) in the limit when $A\rightarrow 0$; we will assume that the process is an adiabatic one. In the limit we first assume for simplicity that we end up with the simple mode expansion
\begin{eqnarray}
\psi_+&\rightarrow&\psi\sim C_{2m}e^{im\varphi}+C_{1m}e^{-im\varphi}\equiv\\
&\equiv&\psi_m+\psi_{-m}\, ,
\end{eqnarray}
where $C_{1m}$ and $C_{2m}$ represent constants and $m$ denotes integers (we suppress for simplicity possible sector dependence in the indices on these constants). Clearly, $m_{\pm}$ in the expressions for $\psi_\pm$ must be identified with $m$. On the assumption that this "spinning down" process is {\it reversible} it follows that we can e.g. choose the $\psi_{\pm}$ states in such a way that (for some $m$)
\begin{equation}
\left\{\begin{array}{l}
\psi_{+}=e^{-\frac{i}{2}A\varphi}\psi_m\, ,\\
\psi_{-}=e^{\frac{i}{2}A\varphi}\psi_{-m}\, .
\end{array}\right.
\end{equation}
Hence,
\begin{equation}
\psi =\psi_++\psi_- \sim e^{-\frac{i}{2}A\varphi}(\psi_m+e^{ iA\varphi}\psi_{-m})\, .
\end{equation}
Clearly, this expression and the logic behind it also works when the simple $\psi_m$-modes are replaced with {\it wave-packets} as they are defined relative to the {\it non-rotating} reference system. The expression for the quantum state $\psi$ in Eq.(71) then coincides (apart from an overall normalization factor) with the expression for the corresponding quantum state which is employed in \cite{Dieks}. The quantum Sagnac phase shift can easily be extracted from Eq.(71) exactly as in \cite{Dieks}.

There is a very interesting point to be made about the form of $\psi_+$ and $\psi_-$ which is connected with the fact that both expressions independently of each other can be used to construct wave-packets. This has an unexpected consequence. In all studies of the classical Sagnac effect as well as of the quantum Sagnac effect {\it counter} rotating beams and particles have been assumed or used. However, let us assume for concreteness that we use the $\psi_+$-states to construct a wave-packet which we write as $e^{-\frac{i}{2}A\varphi}\Phi_+$ while the $\psi_-$-states are used {\it independently} to construct the package $e^{\frac{i}{2}A\varphi}\Phi_-$. Apparently, nothing in this construction, and in particular in the form of the energy eigenvalues, implies that $\Phi_+$ and $\Phi_-$ both can not represent packets which move in {\it either directions} around the rotating disk. Hence, we may consequently construct two wave-packages which both move in the {\it same} direction around the rotating disk but which will give rise to a relative phase-factor which coincides numerically with the relative phase-factor in the quantum Sagnac effect when beams composed of these packages are brought into an interference experiment. This conclusion breaks {\it fundamentally} with the defining construction of the classical Sagnac interferometer and (to the knowledge of this author) all previous theoretical and experimental studies of the quantum Sagnac effect; interfering beams and quantum states are always prepared to move in {\it opposite} directions in space. Quantum mechanics seems to remove this condition entirely. This "peculiar" feature deserves further study elsewhere.

Above we discussed consequences of imposing the fundamental identification on $\Psi_{\pm}$. Let us consider consequences implied by imposing the fundamental identification directly on $\psi_{\pm}$. This can most easily be managed by forcing e.g. $C_{1\pm}=0$ in Eq.(44). Imposing the periodicity condition on the remaining part of the general solution $\psi_+$ implies (a similar expression results from $\psi_-$)
\begin{equation}
\sqrt{A^2+4B}-A\equiv -2p_+\Rightarrow B=-p_+(-p_++A)\, ,
\end{equation}
where $p_+$ denotes integers. The energy spectrum is then given by
\begin{eqnarray}
E_{p_+}&=&\frac{\hbar^2 k^2}{2m_0}+\frac{\hbar^2}{2m_0}(\frac{p_+}{R_0})^2+\hbar p_+\Omega\equiv\nonumber\\
&\equiv& E_{0p_+}+\hbar p_+\Omega\, .
\end{eqnarray}
The structure of the final expression for $E_{p_+}$ is similar to the corresponding expressions in Eq.(48) and Eq.(66) but with the important difference that the second rotation dependent term in Eq.(73) is state and sector dependent and of a genuine quantum mechanical origin. This term describes a {\it quantum angular momentum-rotation coupling}. The angular-momentum eigenvalues are given by
\begin{equation}
\Pi^{\hat{\varphi}}\psi =-\frac{\hbar p_+}{R_0}\psi\, .
\end{equation}
The Lagrangian formulation above demonstrated a formal similarity to a certain level between the system we are considering and the one comprised by a quantum particle coupled to an electromagnetic field. The relation between our system with the fundamental identification imposed on $\psi_\pm$  and the Aharanov-Bohm effect is illuminating. To demonstrate this relation let us endow the quantum particle on the rotating shell with electric charge and direct a magnetic field $\vec{B}$ along the symmetry axis of the shell such that the field is completely confined to the interior of the shell  (ideally at $r=0$) such that the magnetic field is zero on the shell. We may then write
\begin{equation}
A^{\hat{\varphi}}=\frac{\Phi}{2\pi R_0}\, ,
\end{equation}
where $\Phi$ is the total {\it magnetic flux} going through the interior of the shell. Let us also define the following canonical quantities
\begin{equation}
\Phi_L\equiv\frac{2\pi c\hbar}{e}\,\, ,\,\, B_{R}\equiv\frac{\hbar^2}{2m_0R_0^2}\, .
\end{equation}
The energy eigenvalue problem in the {\bf +} sector then becomes
\begin{equation}
(R_0^2\frac{\partial^2}{\partial z^2}+\frac{\partial^2}{\partial\varphi^2}+iA\frac{\partial}{\partial\varphi}+B)\psi_+ =0\, ,
\end{equation}
where
\begin{eqnarray}
&&A\equiv 2\frac{\Phi}{\Phi_L}-\frac{\hbar\Omega}{B_R}\, ,\\
&&B_+\equiv \frac{E_+}{B_R}-(\frac{\Phi}{\Phi_L})^2+\frac{\hbar\Omega}{B_R}(\frac{\Phi}{\Phi_L})\, .
\end{eqnarray}
With the fundamental identification imposed on $\psi_+$ we get, when the state is (for convenience) completely non-localized in the $z$-direction, the following spectrum 
\begin{equation}
E_{p_+}=(p_+-\frac{\Phi}{\Phi_L})(B_R(p_+-\frac{\Phi}{\Phi_L})+\hbar\Omega )\, .
\end{equation}
\begin{figure}[!htpb]
\centering
\includegraphics[]{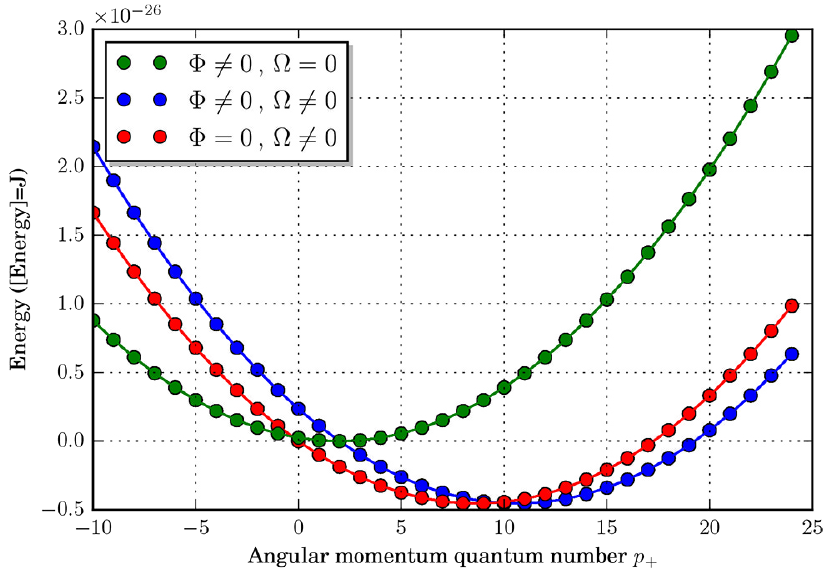}
\caption{Energy spectrum of quantum states with $C_{1+}=0$, $C_{2+}=1$ in Eq.(44) on a rotating cylindrical shell with radius $R_0=10\,\mu m$ and linear velocity $v=R_0\Omega =-100 \,\unitfrac[]{m}{s}$ and with magnetic flux $\Phi =2\Phi_L$ along its symmetry axis as a function of the angular momentum quantum number $p_+$.}
\end{figure}
The spectrum reduces to the one in Eq.(73) when the magnetic flux is set to zero. In Fig.1 we have plotted the energy for a range of $p_+$-values with different combinations of magnetic flux and angular velocity for a micro-sized shell with radius $10 \mu m$ and linear velocity $v=-100  \unitfrac[]{m}{s}$ \footnote{These specific values are chosen by pure convenience.}. We note that $E_{p_+}<0$ for several values on $p_+$ both with and without a magnetic flux present. We observe that these curves exhibit similar shapes as the curve generated by the sole presence of magnetic flux; rotation displaces this curve without changing the overall qualitative features of it. With only magnetic flux present the resulting expression for $E_{p_+}$ and the corresponding plot in Fig.1 reduce to well known ones. It is straightforward to see from our results that a system with both magnetic flux and rotation can be "imitated" exactly on the level of the energy spectrum in {\it certain} cases with only magnetic flux. One example is when we for a fixed value on $p_+$ set
\begin{equation}
\Omega\rightarrow\Omega_{p_+} =\frac{B'_R}{\hbar}(p_+-\dfrac{\Phi}{\Phi_L})
\end{equation}
with {\it some} $B'_R$. We then have that the energy spectrum can be written as if only magnetic flux $\Phi$ and no rotation is present through a shell with radius $R_0'=R_0/\sqrt{2}$. The discussion above goes through ad verbum also for the {\bf -} sector, of course. The expression for $E_{p_+}$ and in particular the plots in Fig.(1) reveal the presence of quantum mechanically induced {\it negative energy states}. Certainly, the presence of such states depends on $\Omega$, and if present, the magnetic flux. It is clear that an {\it arbitrary} number of such states are present provided that the linear velocity $|v|$ is made arbitrary large.

The quantum states we considered in the previous paragraph have some interesting features in relation to the quantum Sagnac effect. However, we will postpone a discussion of these till the next section where we will discuss the quantum mechanics we have constructed in three dimensions. Before we leave the rotating cylinder and this section we emphasize that the presence of negative energy eigenstates seems to represent a generic feature of quantum mechanics in a rotating reference system. They are induced either by a classical state independent correction term to the usual corresponding expressions for the energy in the non-rotating reference system or as a quantum mechanically induced state dependent term. The presence of the negative energy states represents a serious issue for the boost program to define quantum mechanics in a rotating rotating reference system. In \cite{Suzuki} it was proven that the energy of a quantum state generated by a boost transformation is always {\it positive}. It was also argued that this fact corresponds well with a non-relativistic limit of the Dirac equation. At this point we could therefore be led to suspect that the negative energy states reported above represent "artifacts" due to the reduced dimensionality of the theory, even though no assumptions on the dimensionality of space (-time) are made explicitly in the proof in \cite{Suzuki}. However, in the next section we will show that negative energy states are also present in three dimensions.

\section{Quantum Mechanics in A Rotating Cylinder}

In this section we consider the energy eigenvalue problem in the {\it three dimensional} rotating cylinder symmetric coordinate system in Eq.(15). From our earlier elaborations it follows that the energy eigenvalue problem can be formulated as
\begin{equation}
-\dfrac{\hbar^2}{2m_0}\nabla^2\Psi +i\hbar\Omega \dfrac{\partial}{\partial\varphi}\Psi =E\Psi\, .
\end{equation} 
We also have the freedom to act on this equation with ${\cal P}$. However, we will stick to the energy eigenvalue problem in the form of Eq.(82) since it will serve our purposes for now. On general grounds the the Hamiltonian is Hermitian\footnote{See chap. 2 in \cite{Fulling} for a nice and general discussion of the question of Hermicity.}. However, in contrast to the corresponding problem on the rotating shell this equation is not completely separable; not in terms of products or sums or in terms of R-separability. We are consequently not able to provide a general solution to this equation in terms of special functions without having to impose constraints on the energy eigenvalue problem and/or the wave-function $\Psi$. In the previous section we discussed constraints on the wave-function in the dimensionally reduced quantum mechanics. One such constraint was to {\it demand} that the wave-function (i.e. $\psi_\pm$ in Eq.(63)) is periodic in the angular coordinate; i.e., the angular part of the wave-function can be written as a simple exponential $e^{in\varphi}$ where $n$ represents positive and negative integers. This particular ansatz renders the energy eigenvalue problem solvable also in three dimensions. The solutions of the energy eigenvalue problem in Eq.(82) which are regular at the origin are on the form
\begin{equation}
\psi \sim e^{-iEt}e^{in\varphi}e^{ikz}J_n(\sqrt{E^2-n\hbar\Omega}r)\, .
\end{equation}
Let us consider the two most natural boundary conditions for these exact solutions. When we apply the Dirichlet or the Neumann boundary conditions on $\psi$ at $r=R_0$ we get the following expressions for the resulting energy spectra ($D$ denotes Dirichlet, $N$ denotes Neumann)
\begin{eqnarray}
E^{D}_{n,s}&=&\frac{\hbar^2}{2m_0}(\frac{j_{n,s}}{R_0})^2+n\hbar\Omega\, ,\\
E^{N}_{n,s}&=&\frac{\hbar^2}{2m_0}(\frac{j'_{n,s}}{R_0})^2+n\hbar\Omega\, ,
\end{eqnarray}
where $j_{n,s}$ denotes (following the nomenclature in \cite{Watson, Olver}) zero number $s$ of $J_n(x)$. The symbol $j'_{n,s}$ correspondingly denotes zero number $s$ of $J'_n(x)$. The spectra are plotted in Fig.2 for $n=1$, i.e. the lowest non-trivial mode, and its first five zeroes. The macroscopic parameters are the same as for the rotating shell in Fig.(1) with $R_0$ now defining the {\it surface} of the rotating {\it volume}. 

\begin{figure}[!htpb]
\centering
\includegraphics[]{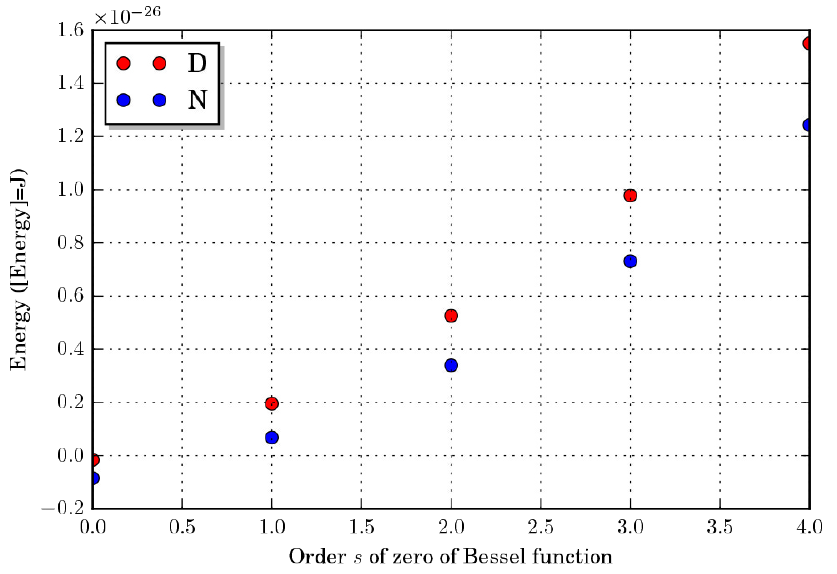}
\caption{Energy spectrum of quantum states with $n=1$ inside a rotating cylinder with radius $R_0=10\mu m$ and linear velocity $v=-100 \unitfrac[]{m}{s}$ with Dirichlet (D) and Neumann (N) boundary conditions imposed at $r=R_0$ as a function of the number of zeroes $s$ of the Bessel function in Eq.(83).}
\end{figure}

We see that the energies are negative for both boundary conditions when $s=0$ with approximate values given by
\begin{equation}
\left\{\begin{array}{l}
E^D_{1,0}\sim -1.5834 \cdot 10^{-28} \mbox{J}\, ,\\
\\
E^N_{1,0}\sim -8.4763\cdot 10^{-28} \mbox{J}\, .
\end{array}
\right.
\end{equation}
This represents a concrete demonstration of the presence of negative energy states also in three space dimensions. Clearly, by letting $R_0$ and/or $\Omega$ increase non-zero $s$-states corresponding to $n=1$ as well as states with higher angular momentum quantum numbers will get shifted downwards and give rise to further negative energy states. We note that the form of the spectra in Eq.(84-85) above is of the same kind as the spectrum in Eq.(73); in both cases the spectra consist of the a term which corresponds to the spectrum in the non-rotating situation corrected with a quantum mechanically induced rotation and state dependent term. This represents a nice illustration of the utility of two (and indeed essentially one-) dimensional model studied earlier in this paper. The other solutions studied there might thus arguably provide a glimpse into the richer but currently theoretically non-derivable quantum physics of the rotating cylinder. The common structure of Eq.(84-85) and Eq.(73) also arguably illustrates the essential one dimensional character of Sagnac effects and these effects as having a {\it kinematical} nature as was pointed out in Sec. 2. 

When we compare our way to tackle the energy eigenvalue problem in three dimensions with our treatment of the corresponding problem with the rotating cylinder in the previous section we immediately acknowledge that we potentially may loose important physics in the three dimensional treatment since the angular momentum quantum number $n$ may encode e.g. a kinematically induced phase. Clearly, no kinematical phase appears in the angular part of $\psi$ in Eq.(83) which otherwise could signal that these states are relative to a moving frame of reference. Hence, the quantum Sagnac effect will apparently not appear in a Sagnac interferometer when states constructed from these states is used, at least not in the form we have encountered the quantum Sagnac effect thus far. We can generalize the ansatz in Eq.(83) slightly by writing the angular part of the wave-function on the form $e^{i\alpha\varphi}$ where $\alpha$ is an arbitrary real number and stil solve the energy eigenvalue problem in terms of Bessel functions \footnote{We note that $\alpha$ can in general be complex-valued. We will keep $\alpha$ real-valued in this work.}. Certainly, this makes for non-periodic wave-functions in the angular direction. This means that we then in principle are also able to construct states which resemble the class ${\bf I}$ states in Eq.(45) by assuming $\alpha =\alpha (m,\Omega )$. States which exhibit similar characteristics are the ones we naturally would expect will be part of the description of states in a typical quantum Sagnac experiment. However, we are unfortunately not able to determine the explicit form of $\alpha (m,\Omega)$ from first principles.

Clearly, the lack of a fundamental understanding of the solution space of Eq.(82) leaves us in darkness as to how the associated quantum state space will look like in general. However, since we do know one exact class of solutions of Eq.(82) (with $\alpha$ an arbitrary real number) there is one avenue worth exploring which might cast more light on its solution space. We might attempt a Lie-theoretic approach and find the Lie-symmetries of Eq.(82) (see e.g. \cite{Olver2}). Knowledge of these symmetries will in general provide us with the means to in principle construct more solutions to Eq.(82). So does the knowledge of the solutions in Eq.(83). The Lie-symmetries of Eq.(82) can in principle be used to map exact solutions into new solutions of the energy eigenvalue problem. This technique to generate new solutions from known ones has been applied in several parts of mathematical physics with great effect (see e.g. \cite{BU}). We note that Eq.(82) is not only central in quantum mechanics in a rotating reference system but also in a non-relativistic limit of electrodynamics in the Born coordinates; an equation with a form which is similar to Eq.(82) emerges as the appropriate equation for describing TM-modes in a resonant rotating micro-cavity \cite{Sunada} \footnote{We note that in that work the authors only consider modes on the form Eq.(83) but not with a general $e^{i\alpha\varphi}$-factor.}. A deeper mathematical understanding of differential equations of the kind represented by eq.(82) is clearly of some importance in both quantum and classical physics. We leave the Lie-theoretical investigation of Eq.(82) to a future publication.

We noted in the previous paragraph that the quantum Sagnac effect in its "ordinary" form will {\it not} be generated by states constructed with the modes in Eq.(83) due to the "regular" part of the angular part of the modes. This holds of course also for the corresponding states on the cylinder in the previous section. There the absence of the "ordinary" quantum Sagnac effect can be easily verified by the approach we developed in Sec. 5. However, it is easily seen that an "anomalous" quantum Sagnac effect is generated by these states. Employing the same approach to calculating the quantum Sagnac phase difference in three dimensions we can employ the states in Eq.(83). Forming a linear superposition $\psi$ of such states with for simplicity the same angular quantum number selected in both sectors and with the Dirichlet condition imposed we get the interference term $\Delta_D\psi$ equal (calculated at $\varphi =0$)
\begin{equation}
\Delta_D\psi \sim 2J^2_n(u_{n,s})\cos (2n\hbar\Omega t)\, ,
\end{equation}
where
\begin{equation}
u_{n,s}\equiv\frac{\hbar}{\sqrt{2m_0}}\frac{j_{n,s}}{R_0}r\, ,
\end{equation}
where we have also used that \cite{Watson}
\begin{equation}
j_{n,s}=j_{-n,s}\, .
\end{equation}
Similar expressions follow when we impose the Neumann condition on both states and when we impose a mixed set of Dirichlet and Neumann conditions on the states. We noted above that Eq.(83) also appears in the study of electromagnetic TM-modes in a rotating system; so will consequently the anomalous Sagnac effect expressed by the above phase differences. This effect was not noted in neither \cite{Sunada} nor any other previous literature known to this author. It might be interesting to investigate whether this novel Sagnac effect might be of technological interest in the context of e.g. laser based gyroscopes.

\section{Discussion}

In this work we have developed a quantum mechanics for a free particle in a non-relativistic rigidly rotating reference system which also can interact with classical electric and magnetic fields. We assumed that the metric structure in the rotating system could be taken to be the usual Euclidean one. We could then directly employ Dirac's quantization procedure in order to construct a quantum mechanics in the rotating system starting with the Lagrangian for a classical point particle. The resulting fundamental commutator algebra was invariant under the geometric reflection transformation ${\cal P}$ while the associated Hamiltonian operator was not. This resulted in two inequivalent Hamiltonians with corresponding non-equivalent states spaces. By solving the associated energy eigenvalue problems it was found that this breaking of the ${\cal P}$-symmetry of the algebra, and of the metric structure, can {\it formally} be considered the {\it source} for the quantum Sagnac effect. The quantum Sagnac effect occurs when two states which do {\it not} belong to the same state space are made to interfere; two quantum states which belong to the {\it same} state space can {\it not} give rise to the quantum Sagnac effect. In contrast to previous treatments of the quantum Sagnac effect our approach emphasizes the intrinsic nature of this effect relative to the rotating reference system. We noted that as a direct consequence of the ${\cal P}$-breaking the quantum Sagnac effect can theoretically be generated by two wave-packets when both packets move in {\it either} direction, at least in the dimensionally reduced theory on the rotating cylinder. The conventional fundamental assumption in dealing with Sagnac effects of any kind is that one directs signals in {\it opposite} directions;  this assumption is thus seemingly proven unnecessary in general. This point seems to warrant further attention both theoretically and experimentally.

We argued that the significance of ${\cal P}$ in the quantum Sagnac effect is directly analogous to the significance of ${\cal P}$ in the classical Sagnac effect within the framework of the special theory of relativity. In this sense the Sagnac effect on the classical level and the quantum Sagnac effect are corresponding phenomena as we would expect them to be on the basis of the very general correspondence principle in quantum mechanics. Our treatment of quantum mechanics in a rotating reference system and the quantum Sagnac effect thus complements the one in \cite{Dieks}; both approaches arguably demonstrate that  non-relativistic quantum mechanics can be seen as a "{\it non-relativistic approximation scheme to a relativistic, Lorentz invariant theory.}" \cite{Dieks}. In \cite{Dieks} the scheme was manifested in the transformation properties of the phase of the wave function under Galilean boost-transformations while in our approach it was manifested in the breaking by the Hamiltonian of a fundamental {\it geometric} reflection symmetry. We provided a brief argument as to how the two approaches are connected. 

Certainly, our formulation of quantum mechanics and the quantum Sagnac effect in rotating coordinates should correspond to results stemming from a non-relativistic limit of relativistic quantum mechanics formulated in the Born-coordinates. However, as we pointed out in the introduction a satisfactory discussion of the Dirac equation in that coordinate system in relation to the quantum Sagnac effect is apparently not currently available in the literature. Furthermore, it was pointed out in \cite{Ryder} that the very formulation of Dirac theory in the case we are considering contains an "ambiguity" due to the special role played by Fermi-Walker transport in defining the local viel-bein. The elaborations in this paper might be of some general utility in the work of shedding more light on this issue. The fundamental role played by ${\cal P}$ in this work should in particular help in guiding the formulation of the Dirac equation in the Born coordinates. Furthermore, Dirac theory is not a consistent theory at higher energies where quantum field theory is the preferred framework. Hence, central aspects of our discussion should be relevant in the context of e.g. QED in a rotating frame of reference and in particular latch naturally to discussions of the kind found in e.g. \cite{Leinaas}.

In Sec. 5 we calculated quantum states on a rotating cylindrical shell. We found that negative energy states appear to be generic on the cylinder. In Sec. 6 we investigated the energy eigenvalue problem in three dimensions. We were able to solve the energy eigenvalue problem exactly for one particular choice of behavior of the wave-function in the angular direction. An analogues configuration also appeared on the cylinder. We showed that these states not only could exhibit states with negative energy but these states could also give rise to an interesting time-dependent interference term; i.e. a constant Sagnac phase shift of the "usual" kind is not generated by this class of states. This last effect has not been discussed in the previous literature on the quantum Sagnac effect. It should in principle be observable in an appropriately designed interference experiment.

Our considerations demonstrate that negative energy eigenstates proliferate in quantum mechanics in a rotating system. This  conclusion runs counter to the conclusion in \cite{Suzuki} that only positive energy states exist in a rotating reference system. That study was based on the boost approach - it thus provides a general proof that the approach to quantum mechanics in rotating coordinates presented in this work supersede the boost approach. So does our exploration of the connection between our formulation of quantum mechanics in a rotating frame and quantum mechanics coupled to electromagnetism in Sec. 5; solutions to the energy eigenvalue problem in the rotating frame are not in general expressible in terms of the solutions to the energy eigenvalue problem in the non-rotating frame multiplied with a simple phase factor.

Do the negative energies calculated in this work have the same physical significance as the corresponding negative energies in the context of e.g. an electron swirling about an atomic nucleus ? Does the value of the lowest energy available to the eigenstates considered, which is bounded from below by the value of the angular velocity, represent the physically true lowest energy available? Clearly, if real the negative energy states should naively give rise to radiative processes. To see this novel possibility we can utilize the idea above of a cylinder spinning up or down but not necessarily in such a way that the quantum states evolve in an adiabatic manner. Focusing on a rotating cylinder with {\it increasing} angular velocity an increasing number of negative energy states are in general expected to condensate out in the bulk with the obvious possibility for spontaneous emission of radiation as particles progressively transition to even lower energy states. This possible phenomenon should in principle be observable as a discrete emission spectrum emanating from the cylinder. Interestingly, this line of thinking can arguably be connected to the phenomenon of {\it superradiance}. 

Certain systems with rotating parts are known to be able to superradiate; e.g. electromagnetic radiation impinged on certain systems triggers rotation dependent energy emission on frequencies corresponding to {\it negative} energy states of the electromagnetic field which is larger than the energy of the incoming radiation on these frequencies. Superradiance thus represents an interesting way to amplify certain electromagnetic signals mechanically. A famous system which exhibits superradiance is a rotating black hole.  There superradiance even has a purely mechanical analog as "energy-mining" through dropping "garbage" into available negative energy states in the ergo-region of the black hole \cite{Misner}, a process known as the Penrose process \cite{Penrose}. Superradiance is also expected to appear in systems containing a rotating cylinder composed of a dielectric material. It was pointed out in \cite{Zeldovich} that the {\it classical} electromagnetic field energy density outside such a cylinder very well can be negative and that superradiance {\it therefore} must occur. This assumption was recently given further weight through a quantum scattering approach \cite{Jaffe}. It is tempting to speculate that the negative energy states uncovered in this work may {\it enhance} such superradiant scattering processes. 

The subject of quantum theory in rotating reference frames is a highly fragmented one and various approaches to quantum mechanics and quantum field theory in rotating frames are found in the literature. The approaches to quantum mechanics in rotating reference frames all arguably suffer from lack of "rigor". In this paper we have attempted to give that subject a firmer and more coherent theoretical basis. As a prerequisite for such a formulation we provided a reanalysis of the classical Sagnac effect in order to pin-point exactly which feature of the classical theory which triggers the classical effect; the same feature is arguably also the source for the quantum Sagnac effect. This quantum effect is important since it represents one of the few indisputable {\it empirical} evidences for effects arising in quantum physics in accelerated reference frames. A quantum mechanics in a rotating reference system must be able to accommodate this effect. In the process of devising a coherent quantum mechanics which sits well with well established theoretical frameworks in physics and with experimental findings several novel insights about quantum mechanics in a rotating frame have been gained in this work, some of which can be tested experimentally. We have also pointed to several directions for further theoretical and experimental study. Interestingly, some are also of importance to classical electrodynamics and the Sagnac effect in rotating micro-devices. We hope to return to these questions in the future.


\begin{thebibliography}{lll}

\bibitem{Leinaas} J.I. Korsbakken, J.M. Leinaas, Phys.Rev. D{\bf 70} 084016 (2004).

\bibitem{Rauch} H. Rauch, S.A. Werner, {\it Neutron Interferometry, Lessons in Experimental Quantum Mechanics, Wave-Particle Duality, and Entanglement}, Oxford University Press (2015).

\bibitem{Sagnac} M.G. Sagnac, Comptes rendus de l'Académie des sciences {\bf 157} 708 (1913).

\bibitem{Werner} S.A. Werner, J.L. Staudenmann, R. Colella, Phys. Rev. Lett. {\bf 42} 1102 (1979).

\bibitem{Anandan_term} J. Anandan, Phys. Rev. D{\bf 24} 338 (1981).

\bibitem{Macek} W.M. Macek, D.T.M. Davis, Jr., Appl. Phys. Lett. {\bf 2} 67 (1963).

\bibitem{Anderson} R. Anderson, H.R. Bilger, G.E. Stedman, Am. J. Phys. {\bf 62} 975 (1994).

%\bibitem{WHO?} WHO?, {\it Quantum Mechanics in a Rotating Frame} in {\it Relativity in Rotating Frames}, ........ (Editor), Kluwer Academic Publ. (2004).

\bibitem{Search} C. P. Search, J. R. E. Toland, M. Zivkovic, Phys. Rev. A{\bf 79} 053607 (2009).

\bibitem{Mashhon} B. Mashhon, Phys. Rev. Lett. {\bf 61} 2639 (1988).

\bibitem{Suzuki} J. Anandan, J. Suzuki, {\it Quantum Mechanics in a Rotating Frame} in {\it Relativity in Rotating Frames, Relativistic Physics in Rotating Reference Frames}, G. Rizzi, M.L. Ruggiero (Editors), Kluwer Academic Publ. (2004).

\bibitem{AharonovC} Y. Aharanov, G. Carmi, Found. Phys. {\bf 3} 493 (1973).

\bibitem{Aharonov} Y. Aharonov, A. Casher, Phys. Rev. Lett. {\bf 53} 319 (1984).

\bibitem{Sakurai_1980} J.J. Sakurai, Phys. Rev. D{7bf 21} 2993 (1980).

\bibitem{Hendriks} B.H.W. Hendriks, G. Nienhuis, Quantum Opt. {\bf 2} 13 (1990).

\bibitem{Mashhon_2} B. Mashhon, Phys. Rev. Lett. {\bf 68} 3812 (1992).

\bibitem{Anandan} J. Anandan, Phys. Rev. Lett. {\bf 68} 3809 (1992). 

\bibitem{Rizzi} G. Rizzi, M.L. Ruggiero {\it The relativistic Sagnac effect: two derivation} in {\it Relativity in Rotating Frames, Relativistic Physics in Rotating Reference Frames}, G. Rizzi, M.L. Ruggiero (Editors), Kluwer Academic Publ. (2004).

\bibitem{Lord} Lord Rayleigh, Proc. London Math. Soc. {\bf 17} 4 (1885).

\bibitem{Bryan} G.H. Bryan, Proc. Cambridge Phil. Soc. {\bf VII} 101 (1890).

\bibitem{Xia} D. Xia, C. Yu, L. Kong, Sensors {\bf 14} 1394 (2014).

\bibitem{Yu} Y. Yu et. al., Micromachines {\bf 8 (1)} 2 (2017).

\bibitem{Dieks} D. Dieks, G. Nienhuis, Am. J. Phys. {\bf 58} 650 (1990).

\bibitem{Dieks2} D. Dieks, Eur. J. Phys. {\bf 12} 253 (1991).

\bibitem{Langevin} P. Langevin, Comptes rendus {\bf 200} 48 (1935).

\bibitem{Rosen} N. Rosen, Phys. Rev. {\bf 71} 54 (1947).

\bibitem{Arditty} H.J. Arditty, H.C. Lefevre, Optics Letters {\bf 6} (8) 401 (1981) and in {\it Theoretical Basis of Sagnac Effect in Fiber Gyroscopes} in {\it Fiber-Optic Rotation Sensors and Related Technologies, Proc. of the First Int. Conf. MIT, Cambr., Mass., USA, November 9-11, 1981}, Springer-Verlag (1982).

\bibitem{Vugalter} G. A. Vugal'ter, G. B. Malykin, Radiophysics and Quantum Electronics {\bf 42} (4) 333 (1999).

\bibitem{Wang} R. Wang, Y. Zheng, A. Yao, D. Langley, Phys. Lett. A{7bf 312} 7 (2003); R. Wang, Y. Zheng, A. Yao, Phys. Rev. Lett. {\bf 93} 143901 (2004).

\bibitem{Angelo} A. Tartaglia, M. L. Ruggiero, Am. J. Phys. {\bf 83} 427 (2015).

\bibitem{Rizzi_Serafini} G. Rizzi, A. Serafini {\it Synchronization and desynchronization on rotating platforms} in {\it Relativity in Rotating Frames, Relativistic Physics in Rotating Reference Frames}, G. Rizzi, M.L. Ruggiero (Editors), Kluwer Academic Publ. (2004).

%\bibitem{Kuzera} B. Jensen, J. Kucera, J. Math. Phys. {\bf 34} (11) 4975 (1993).

\bibitem{AB} Y. Aharonov, D. Bohm, Phys. Rev. {\bf 115} 485 (1959).

\bibitem{Dirac} P.A.M. Dirac, {\it Lectures on Quantum Mechanics}, Belfer Graduate School, Yeshiva University, New York (1964).

\bibitem{Landau} L.D. Landau, E.M. Lifshitz, {\it Mechanics, Course of Theoretical Physics, Volume 1}, Pergamon Press (1976).

\bibitem{Bergmann} J. L. Anderson, P. G. Bergmann, Phys. Rev.(2) {\bf 83} 1018 (1951).

\bibitem{Teitelboim} M. Henneaux, C. Teitelboim, {\it Quantization of Gauge Systems}, Princeton University Press (1992).

\bibitem{Leaf} B. Leaf, Am. J. Phys. {\bf 47} 811 (1979).

\bibitem{Gruber} G.R. Gruber, Found. Phys. {\bf 1} 227 (1971).

\bibitem{daCosta} H. Jensen, H. Koppe, Ann. Phys. (N.Y.) {\bf 63} 586 (1971), R.C.T. da Costa, Phys. Rev. A{\bf 23} 1982 (1981).

\bibitem{Jensen} B. Jensen, R. Dandoloff, Phys. Lett. A{\bf 375} 448 (2011), B. Jensen, R. Dandoloff, A. Saxena, Physica Scripta {\bf 89} 105202 (2014).

\bibitem{Sakurai} J.J. Sakurai, {\it Advanced Quantum Mechanics}, Benjamin/Cummings Publishing Company (1984).

\bibitem{Tinkham} M. Tinkham, {\it Introduction to superconductivity}, Dover Publications (1996).

\bibitem{Anandan_1977} J. Anandan, Phys. Rev. D{\bf 15} 1448 (1977).

\bibitem{Fulling} S.A. Fulling, {\it Aspects of Quantum Field Theory in Curved Space}, Cambridge University Press (1989).

\bibitem{Watson} G.N. Watson, {\it A treatise on the theory of Bessel functions}, Cambridge University Press (1952).

\bibitem{Olver} F.W.J. Olver, {\it Bessel functions, Part III, Zeros and associated values}, Cambridge University Press (1960).

\bibitem{Olver2} H. Stephani, {\it Differential equations: their solution using symmetries}, edited by M. MacCallum, Cambridge University Press (1989).

\bibitem{BU} B. Jensen, U. Lindstr\"{o}m, Phys. Rev. D{\bf 52} 3543 (1995).

\bibitem{Sunada} S. Sunada, T. Harayama, Phys. Rev. A{\bf 74} 021801(R) (2006).

\bibitem{Ryder} L. Ryder, Gen. Rel. Gravit. {\bf 40} 1111 (2008).

\bibitem{Misner} C.W. Misner, K.S. Thorne, J.A. Wheeler, {\it Gravitation}, W.H. Freeman and Company (1973).

\bibitem{Penrose} R. Penrose, R.M. Floyd, Nature Physical Science {\bf 229} 177 (1971).

\bibitem{Zeldovich} Y.B. Zel'dovich, L.V. Rozhanskii, A.A. Starobinskii, Radiophys. Quantum Electron. {\bf 29} 761 (1986).

\bibitem{Jaffe} M.F. Maghrebi, R. L. Jaffe, M. Kardar, Phys. Rev. Lett. {\bf 108} 230403 (2012).

\end{thebibliography}
\end{document}